\documentstyle[12pt]{article}
\begin{document}
\begin{titlepage}

\centerline{\large \bf Violation of Finite-Size Scaling in Three Dimensions} 
\vspace*{0.4cm}
\centerline{X.S. Chen$^{1,2,a}$ 
and V. Dohm$^{1,b}$}
\centerline{$^1$ Institut f\"{u}r Theoretische Physik, Technische Hochschule 
Aachen,}
\centerline{D-52056 Aachen, Germany}
\centerline{$^2$ Institute of Particle Physics, Hua-Zhong 
Normal University,}
\centerline{Wuhan 430079, P.R. China}
\vspace*{0.3cm}

\begin{abstract}
We reexamine the range of validity of finite-size scaling in the  
$\varphi^4$  lattice model and the $\varphi^4$ field theory below four 
dimensions.
We show that general renormalization-group arguments based on the 
renormalizability of the $\varphi^4$ theory 
do not rule 
out the possibility of a violation  of finite-size scaling due to a finite
lattice constant  and a finite cutoff. 
For a confined geometry of linear size $L$ with periodic boundary 
conditions we analyze the approach towards bulk critical behavior as 
$L \to \infty$ at fixed $\xi$ for $T > T_c$ where $\xi$ is the bulk
correlation length. We show that for this analysis ordinary renormalized 
perturbation theory is sufficient. On the basis of one-loop results 
and of exact results in the spherical limit we find that finite-size 
scaling is violated for both 
the $\varphi^4$ lattice model  and the $\varphi^4$ field theory  
in the  region $L \gg \xi$. The non-scaling effects in the field theory 
and in the lattice model differ significantly  from each other. 
\end{abstract}


$^a$ e-mail: chen@physik.rwth-aachen.de  
\newline
$^b$ e-mail: vdohm@physik.rwth-aachen.de
\end{titlepage}

\section{ Introduction}

One of the fundamental achievements of the renormalization-group (RG)
theory of critical phenomena is the elucidation and proof of universality and
scaling near critical points [1-3]. 
These predictions have been shown to be asymptotically exact sufficiently
close to criticality of infinitely large systems. For finite or partially
finite systems, the field-theoretic version of RG theory has also provided
an apparently exact prediction of universal finite-size scaling for systems
with periodic boundary conditions \cite{6}, in accord with phenomenological
considerations \cite{1} and with numerous analytical and numerical studies
in statistical and elementary particle physics in the past decades [3, 6-12]. 
Thus the validity of finite-size scaling appears to be well established.

Consider, for example, the susceptibility 
$\chi(t,L)$ of a ferromagnetic system for $t = (T - T_c)/T_c \geq 0$ 
in a $d$-dimensional finite geometry with a characteristic size $L$. 
For large $L$ and small $t$
the property of finite-size scaling means
that $\chi$ has the asymptotic form
\begin{equation}
\chi(t,L) = \chi (t, \infty) f (L/\xi)
\end{equation}
where $\chi (t, \infty) = A_\chi t^{- \gamma}$ is the bulk susceptibility
and $\xi = \xi_0 t^{-\nu}$ is the bulk correlation length. 
For a given geometry and periodic boundary conditions the scaling function 
$f(x)$ was 
found \cite{6} to be universal 
for $d < 4$ 
which implied that the relative deviation from bulk critical behavior
\begin{equation}
\Delta \chi \equiv \frac{\chi (t, \infty) - \chi (t , L)} {\chi (t, \infty)} 
= g(L/\xi)
\end{equation}
is  universal as well in the entire range $0 \leq L/\xi  \leq \infty$ 
with $g (\infty) \; = \; 1 - f(\infty) \; = \; 0$. This result, if correct,
agrees with the scaling hypothesis \cite{1} which implies 
that the approach
towards bulk critical behavior ($L/\xi \to \infty$ at fixed $\xi < \infty$
above $T_c$) can be embodied in the scaling function  $g(L/\xi)$ \cite{1}.  
Universality in this context means that the shape of the scaling function 
$g(x)$ depends on the
geometry and on the boundary conditions but does not depend on any nonuniversal
parameter, in particular not on the lattice constant $\tilde{a}$ of lattice
models or on the cutoff $\Lambda$ of field theories. As a 
consequence it was generally accepted that finite-size scaling functions such
as $g(x)$ can be calculated  on the basis of
field theories in the limit $\Lambda \to \infty$ [3,4,7-50].

Br\'{e}zin's RG analysis \cite{6} started from the $\varphi^4$ lattice
model with a finite lattice spacing $\tilde{a}$. The RG arguments,
however, were
presented within the renormalized theory after the limit $\tilde{a} \to 0$
was taken. This limit is usually considered  
in studies of renormalized field theory 
of {\it{bulk}} systems for $d < 4$ [3, 13, 51]
where cutoff and lattice effects are known to yield only subleading 
corrections to the 
leading critical temperature dependence. The asymptotic
unimportance of cutoff and lattice effects 
also for {\it{confined}} systems appeared to be
a plausible assumption that was not questioned in Ref. \cite{6} but is 
checked in the present paper and is found to be invalid.

Very recently we have shown \cite{15} that this latter assumption is not 
generally justified in the O($n$) symmetric 
$\varphi^4$ {\it field} theory of confined systems with periodic boundary
conditions. Specifically it was shown in the large-$n$ limit for 
$2 < d < 4$ that a finite 
cutoff $\Lambda$ implies 
a violation of 
finite-size scaling in the region $L/\xi \gg1$ above $T_c$, 
with a non-exponential and nonuniversal approach 
$\Delta \chi \propto (\Lambda L)^{-2}$ 
towards zero, even arbitrarily close to $T_c$. 
This behavior was traced back to the  $(\nabla \varphi)^2$ term
in the field-theoretic $\varphi^4$ Hamiltonian which approximates the more
general interaction $(\varphi_i - \varphi_j)^2$ of the $\varphi^4$ lattice
model. In the latter model an exponential size dependence in the
region $L/\xi \gg 1$ was found \cite{15}, 
in accord with previous 
results for exactly solvable model systems [5,6,14, 53-57]. 
The possibility of a violation of finite-size scaling  in the $\varphi^4$ 
{\it lattice} model at  finite lattice spacing, however, was not yet analyzed 
in our recent work  \cite{15}. Thus the important question remained open 
whether
the violation of finite-size scaling found in the $\varphi^4$ field theory
\cite{15}
is an artifact of the field-theoretic 
continuum approximation or 
whether finite-size scaling breaks down more generally for 
$L/\xi \gg 1$ in confined  lattice systems with a finite lattice constant.

It is the purpose of the present paper to take up this problem for the  O$(n)$ 
symmetric $\varphi^4$ lattice model at finite lattice constant $\tilde{a}$
in the context of a detailed RG analysis,  without taking the limit 
$\tilde{a} \to 0$. We assume renormalizability in terms of bulk 
renormalizations and thus work for 
dimensionality $d$ below the upper critical
dimension which is $4$ in our case. Thus this may become relevant to real 
three-dimensional systems. We shall show that the 
renormalizability of the $\varphi^4$ model in a confined geometry  implies
the asymptotic $(L \to \infty, \xi \to \infty)$ validity of finite-size
scaling  for $d < 4$ at any {\it{fixed finite}} 
ratio $L/\xi < \infty$, in agreement 
with the proof of 
Br\'{e}zin \cite{6},
but does not rule out a violation of finite-size scaling in the limit
$L/\xi \to \infty$ at finite $\tilde{a}/\xi > 0$. 
On the basis of one-loop results for general $n$
and of exact results in the large-$n$ limit we indeed find such a violation: 
instead of (2) the more general form
\begin{equation}
\Delta \chi \; = \; g(L/\xi) \; 
\left[ 1 \; + \; R(L/\xi, \; \tilde{a}/\xi) \right]
\end{equation}
must be considered where $g(x)$ is universal but where the nonuniversal
function $R$ contains a nontrivial dependence on the lattice
constant $\tilde{a}$. Although $R$ vanishes for fixed finite ratio 
$L/\xi < \infty$ in the asymptotic region,
\begin{equation}
\lim_{(L\; ,\;\xi) \; \to \infty} 
\; R(L/\xi, \tilde{a}/\xi)\; = \; R (L/\xi, 0) 
\; = \; 0 \; , \quad  L/\xi \; \;{\mbox{fixed}}, 
\quad 
\; L/\xi < \infty \quad ,
\end{equation}
it exhibits a singular behavior in approaching the bulk limit 
$L/\xi \to \infty$ at any fixed $\tilde{a}/\xi > 0$,
\begin{equation}
\lim_{x \to \infty} \; R (x, \tilde{a}/\xi) \; = \; \infty \; , \quad 
\tilde{a}/\xi 
\; {\mbox{fixed}}, \quad  
\tilde{a}/\xi > 0 \quad .
\end{equation}
This implies that for sufficiently large $L/\xi$ the leading size
dependence
\begin{equation}
\Delta \chi \; \sim \; g (L/\xi) \; R(L/\xi, \tilde{a}/\xi) \quad , \quad
L/\xi \gg 1 \quad , \; \tilde{a}/\xi > 0,
\end{equation}
is nonuniversal and violates finite-size scaling.
We emphasize that this violation is not a subleading non-asymptotic property
but occurs in leading order at any finite $\xi < \infty$, even arbitrarily
close to $T_c$ where ''corrections to scaling''  or corrections of {\it{bulk}}
properties due to a finite lattice constant are completely negligible.
(Here and in the following the symbol $\sim$ means asymptotic behavior 
including the amplitude, i.e., $\Delta \chi \sim G(x)$ for $ x \gg 1$ means 
$\lim_{x \rightarrow \infty} \Delta \chi/G(x) = 1$.)

Our results imply that the property of finite-size scaling (for confined
systems with periodic boundary conditions) which was previously believed
to be exact  for $d < 4$ in the asymptotic 
($L \gg \tilde{a}, \xi \gg \tilde{a}$) region of the 
$L^{-1} - \xi^{-1}$ plane (Fig. 1) is
not valid in a small but important part of this region 
(below the dashed line in Fig. 1).
The nonuniversal
function $R$ is negligible at $T = T_c$ for sufficiently large $L$ 
but it increases as $L/\xi$ increases above $T_c$ 
at fixed $\tilde{a}/\xi > 0$. 
The approach to  the bulk limit
(arrow in Fig. 1) corresponds to a crossover from the scaling region to a 
non-scaling region where nonuniversal effects due to the finite lattice
constant dominate the finite-size deviations from bulk behavior. 
The location of the (smooth)
crossover region may be characterized by the line along which 
$R (L/\xi, \tilde{a}/\xi) \; \simeq \; 1$, i.e., where the scaling and
non-scaling contributions to $\Delta \chi$ are equally large. 
This requirement defines
the dashed line in Fig. 1. A similar line should exist below $T_c$. 
Explicit results for $R$ will be given in Sections 4 and 5 for the
$\varphi^4$ lattice model for $2 < d \leq 4$. 

Essential features of these results will remain valid even for $d > 4$ 
\cite{CDnew} such that the finite-size scaling form of Privman and Fisher 
\cite{PFi,VP} will be violated for $L \gg \xi$.
In particular, the lowest-mode approach of Br\'ezin and Zinn-Justin \cite{7} 
and the phenomenological single-variable scaling form of Binder et al. 
\cite{BNPY,KB85} fail qualitatively for 
$L \gg \xi \gg \tilde{a}$ where these theories predict a 
{\it universal power-law} behavior $\Delta \chi \propto L^{-d}$ above four 
dimensions, rather than a {\it non-universal exponential} behavior 
\cite{15,CDnew}
$\Delta \chi \propto e^{-c L}$ as derived in Sections 4 and 5 of the present
paper. Such striking {\it structural} differences \cite{ChDo3} between the 
lowest-mode approximation and the effects of the higher modes cannot be 
regarded only as "corrections" \cite{LBB}.

For comparison we also calculate the function $R_{field} (L/\xi, \Lambda \xi)$
for the field-theoretic $\varphi^4$ model at finite cutoff $\Lambda$
(with periodic boundary conditions).
For $d < 4$ we find a violation of finite-size scaling \cite{15}
due to a divergence
of $R_{field} (L/\xi, \Lambda \xi)$ in the limit $L/\xi \to \infty$ at
fixed $\Lambda \xi < \infty$, analogous to (5) for the lattice model. 
The form of $R_{field}$ of the $\varphi^4$ field
theory, however, differs significantly from that of $R$ for the 
$\varphi^4$ lattice model. Even the {\it{sign}} of $R_{field} < 0$
is different from that of $R > 0$. Thus the $\varphi^4$ field theory 
based on the standard Landau-Ginzburg-Wilson continuum Hamiltonian
does not predict the correct structure of the leading finite-size 
deviation from bulk critical behavior of lattice systems at any $T > T_c$ 
(and presumably at any $T < T_c$) for $d < 4$. 
We show that this statement remains
valid also for $d = 4$ which may be relevant to elementary particle physics
\cite{GL92,MO96,PH}, to disordered systems \cite{Spain}, and more generally 
to systems at their upper critical dimension \cite{Fischer1}. For $d > 4$ 
the failure of the continuum approximation is even more severe as it pertains 
to the entire $L^{-1} - \xi^{-1}$ plane [58,63,66-68].

From a purely quantitative point of view, the non-scaling behavior of
$\chi$ is a small effect that occurs  
predominantly in a region where the total finite-size contributions are
exponentially small (for periodic boundary conditions). From a more 
fundamental point of view, however, the
violation of finite-size scaling below four dimensions is a matter of 
principle, regardless how 
small this effect might be. In particular, our RG analysis for the simplest 
case of {\it{periodic}} boundary
conditions  raises considerable doubt about the validity of 
finite-size scaling  in the more complicated cases of {\it{non-periodic}}
boundary conditions where additional renormalizations and nonuniversal 
length scales come into play. They imply that general RG arguments
are less compelling since additional assumptions would be needed. Furthermore, 
possible non-scaling effects for non-periodic boundary conditions may no 
longer be exponentially small. This may 
open up the prospect of resolving the longstanding and recent problems 
concerning the interpretation \cite{Dohm1} of experimental
data for confined $^4$He near the superfluid transition [69-80]. Work
in this direction is in progress.

Our paper is organized as follows. In Section 2 we describe the 
renormalization scheme for the bulk $\varphi^4$
lattice model at finite lattice spacing. In Section 3 we extend this scheme
to the confined system and 
show that the RG arguments do not rule out a violation of
finite-sized scaling in the limit $L \gg \xi$.
In Section 4 we calculate $\Delta \chi$ for $d \leq 4$  in one-loop order
for general $n$. The exact result for
this quantity in the large-$n$ limit is derived in Section 5. 
A summary and further discussion of our results  is given in Section 6.

\section{Lattice Model: Bulk Properties at Finite Lattice Constant}

In this Section we introduce our notation and define the renormalization of
bulk quantities of the $\varphi^4$ lattice model at finite lattice spacing
above $T_c$. This will serve as the framework for the renormalization-group 
analysis of finite-size effects at finite lattice spacing in the subsequent 
Sections.
A detailed formulation of the theory at finite lattice spacing is
indispensable for  clearly distinguishing lattice effects of the
finite system from ordinary non-asymptotic Wegner \cite{Wegner} corrections 
to scaling.

We consider a $\varphi^4$ lattice Hamiltonian $H$ for the 
variables $\varphi_i$ 
on the lattice points ${\bf{x}}_i$ of a simple-cubic lattice in a cube with 
volume $V = L^d$ and with periodic boundary conditions. (Generalizations to
different geometries will be considered in the subsequent Sections.) We assume
the statistical weight $\propto e^{-H}$ with
\begin{equation}
H = \tilde{a}^d \left\{ \sum\limits_i \left[ \frac{r_0}{2} \varphi_i^2  +
u_0(\varphi^2_i)^2 \right] + \sum\limits_{i,j} \frac{1}{2 \tilde{a}^2}
J_{ij} (\varphi_i - \varphi_j)^2 \right\} 
\end{equation} 
where $\tilde{a}$ is the lattice constant. The variables $\varphi_i$
have $n$ components 
$\varphi_{i \alpha}$ with 
$\alpha = 1,2,\cdots,n $ which vary in the range 
$ - \infty \leq \varphi_{i \alpha} \leq \infty$. 
The couplings $J_{ij}$ are dimensionless quantities whereas the variables
$\varphi_i$ have the dimension $[\tilde{a}^{(2-d)/2}]$ and $\tilde{a}$ has 
the dimension of a length.

\subsection{Unrenormalized Theory}

The renormalization-group treatment of this
model in the bulk limit $V \to \infty$ is well known
which is usually carried out
in the limit of zero lattice spacing $\tilde{a} \to 0$ 
or in the limit of infinite cutoff $\Lambda \to \infty$ in the continuum 
version (see (54) below) \cite{14,Amit,13}.
Here we shall formulate the renormalization of bulk quantities at finite
$\tilde{a}$. Similar to the previous formulation of the bulk theory at
fixed $d < 4 \; $  \cite{SD} we shall express the bare theory in 
terms of the bulk
correlation length $\xi$ before turning to the renormalized theory.

We start from the bulk two-point vertex function \cite{14,Amit,13}
\begin{equation}
\Gamma^{(2)} ({\bf{k}}, r_0, u_0, \tilde{a}, d) \quad = 
\quad \chi_b ({\bf{k}})^{-1}
\end{equation}
where $\chi_b({\bf{k}})$ is the bulk susceptibility at finite wavevector 
${\bf{k}}$ above
$T_c$
\begin{equation}
\chi_b({\bf{k}}) \; = \; \lim\limits_{L \to \infty} 
\frac{\tilde{a}^{2d}}{L^d}
\sum\limits_{i,j} < \varphi_i \varphi_j > 
e^{-i{\bf{k}} \cdot ({\bf{x}}_i - {\bf{x}}_j)}\quad .
\end{equation}
It serves to define the bulk correlation length $\xi$ above $T_c$ according
to
\begin{equation}
\xi^2 \; = \; \chi_b({\bf{0}}) \frac{\partial}{\partial k^2} 
\left[ \chi_b({\bf{k}})  \right]^{-1} \; _{\Big | {\bf{k}} = {\bf{0}}} \quad  .
\end{equation}
We shall also consider the four-point vertex function $\Gamma^{(4)}$ of the
bulk theory \cite{14,Amit,13} at vanishing external wavenumber.

The parameter $r_0$ in $H$ is taken to be a linear function of the reduced
temperature
\begin{equation}
t \; = \; (T - T_c)/T_c , \quad r_0 \; = \; r_{0c} + a_0 t \quad , 
\end{equation}
with $a_0 > 0$. The critical value $r_{0c}$ of $r_0$ is determined by 
$\chi_b ({\bf{0}})^{-1} = 0$, i.e.
\begin{equation}
\Gamma^{(2)} ({\bf{0}}, r_{0c}, u_0, \tilde{a}, d) = 0 \quad ,
\end{equation}
which provides an implicit definition of the function
\begin{equation}
r_{0c} = r_{0c}(u_0, \tilde{a}, d)
\end{equation}
at finite lattice spacing $\tilde{a}$. (Note that $r_{0c}(u_0, \tilde{a}, d)$ 
does not have an expansion in integer powers of $u_0$, unlike the vertex 
functions $\Gamma^{(N)} (r_0, u_0, \tilde{a}, d)$ 
\cite{SD}.)

Instead of $r_0$ we shall substitute
\begin{equation}
r_0 = r_0 - r_{0c} \; + \; r_{0c}  (u_0, \tilde{a}, d)
\end{equation}
into $\Gamma^{(N)}$ and consider  $\Gamma^{(N)}$ (at ${\bf{k}} = 
{\bf{0}})$ as well
as the correlation length $\xi$ as functions of 
$r_0 - r_{0c}, u_0, \tilde{a},d,$
\begin{equation}
\Gamma^{(N)} \; = \; \Gamma^{(N)} (r_0 - r_{0c}, u_0, \tilde{a},d) \; ,
\end{equation}
\begin{equation}
\xi \;  = \;  \xi\;  (r_0 - r_{0c}, u_0, \tilde{a},d) \; .
\end{equation}
Since $\xi$ is a monotonic function of $r_0 - r_{0c}$, equation (16) can be 
inverted to define $r_0 - r_{0c}$ as a function of $\xi$,
\begin{equation}
r_0 - r_{0c} \; = \; h (\xi, u_0, \tilde{a}, d) \; .
\end{equation}
Finally we may substitute (17) into (15) which leads to bare vertex functions
$\tilde{\Gamma}^{(N)}$ at finite $\tilde{a}$ in terms of $\xi$,
\begin{equation}
\tilde{\Gamma}^{(N)} (\xi, u_0, \tilde{a}, d) \; = \; 
\Gamma^{(N)}(h (\xi, u_0, \tilde{a}, d), u_0, \tilde{a}, d) \quad .
\end{equation}
These definitions are parallel to those at infinite cutoff in 
Ref. \cite{SD}. (In particular, the functions $\tilde{\Gamma}^{(N)}$ have
an expansion in integer powers of $u_0$.) 
We illustrate these definitions by the one-loop results
\begin{eqnarray}
h(\xi, u_0, \tilde{a},d) \; & = & \; J_0 \xi^{-2} 
\left\{ 1 + 4(n+2) u_0 \int\limits_{\bf{k}} 
\left[ \hat{J}_{\bf{k}} (  \hat{J}_{\bf{k}} \; + \;
J_0 \xi^{-2} )\right] ^{-1} 
+ O(u^2_0) \right\}\; , \quad \quad  \\[1cm]
\tilde{\Gamma}^{(2)} ({\bf{k}}, \xi, u_0, \tilde{a}, d) \; 
& = & \; \hat{J}_{\bf{k}} + J_0 \xi^{-2} + O (u^2_0) \; \qquad ,\\[1cm]
\tilde{\Gamma}^{(4)} (\xi, u_0, \tilde{a}, d) \; & = & \; 24 u_0 \left\{
1 - 4 (n + 8) u_0 \int\limits_{\bf{k}}  
\left[ \hat{J}_{\bf{k}} \; + \; J_0 \xi^{-2} \right]^{-2} \; + O (u^2_0)
\right\}\qquad
\end{eqnarray}
where
\begin{equation}
\hat{J}_{\bf{k}}\; = \; \frac{2}{\tilde{a}^2}\left[J(0) - J({\bf{k}})\right]
\end{equation}
with
\begin{equation}
J({\bf{k}}) \; = \; (\tilde{a}/L)^d \; \sum\limits_{i,j} \; J_{ij}
e^{-i{\bf{k}} \cdot ({\bf{x}}_i - {\bf{x}}_j)} \quad .
\end{equation}
We assume a finite-range pair interaction $J_{ij}$ such that its Fourier
transform has the small ${\bf{k}}$ behavior
\begin{equation}
\hat{J}_{\bf{k}}\; = \; J_0 {\bf{k}}^2 \; + \; O(k^2_i k^2_j)
\end{equation}
with a finite constant
\begin{equation}
J_0 \; = \; \frac{1}{d} 
(\tilde{a}/L)^d \; \sum\limits_{i,j} (J_{ij}/\tilde{a}^2)
({\bf{x}}_i - {\bf{x}}_j)^2 \; > 0 \quad .
\end{equation}
The dependence of the quantities (19) - (21) on $\tilde{a}$ comes from
the integration limit of the bulk integral
\begin{equation}
\int\limits_{\bf{k}} \; \equiv \; \int \; \frac{d^dk}{(2\pi)^d}
\end{equation}
with $|k_j| \leq \pi / \tilde{a} \; , \; j = 1, 2, \cdots, d$.
Because of the super-renormalizability of the $\varphi^4$ theory 
\cite{14,Amit,13}
the bare functions $\tilde{\Gamma}^{(N)}$ remain finite in the limit
$\tilde{a} \to 0$ at fixed $\xi$ and $u_0$ for $d < 4$.

\subsection{Renormalization at Finite Lattice Constant}

As is well known, the perturbative results (19) - (21) of the bare theory 
do not provide a correct description in the critical region 
$\xi \gg \tilde{a}$ for $d \leq 4$. This problem is
circumvented by turning to the renormalized theory which provides a
mapping from the critical to the non-critical region where perturbation
theory is applicable [see (44) below]. We start from the bare $N$-point 
vertex functions $\tilde{\Gamma}^{(N)}$
as functions of the correlation length $\xi$ where $\xi$ is considered to
be a given quantity. 
The explicit
determination of $\xi$ as a function of the reduced temperature $t$ 
at finite $\tilde{a}$ is a 
separate issue \cite{SD} that is postponed to Appendix A.

Since $\xi$ does not require a renormalization it suffices to introduce only
two renormalization factors $Z_\varphi$ and $Z_u$ to define multiplicatively
renormalizable vertex functions $\tilde{\Gamma}^{(N)}_R$. 
We define the renormalized variable
\begin{equation}
\varphi^R_i \; = \; Z_\varphi^{-1/2} \varphi_i
\end{equation}
and the renormalized coupling
\begin{equation}
u \; = \; J^{-2}_0\; \mu^{-\epsilon}\; \; Z^{-1}_u  \; u_0
\end{equation}
with $\epsilon = 4-d$.
The reference length $\mu^{-1}$ is arbitrary. (It can be conveniently
chosen as $\mu^{-1} = \xi_0$ where $\xi_0$ is the amplitude of the asymptotic 
bulk correlation length as specified in (A.17) of  Appendix A.)
The definitions (27) and (28) lead to the following renormalized vertex
function $\tilde{\Gamma}_R^{(2)}$ (at finite ${\bf{k}}$) and   
$\tilde{\Gamma}_{R}^{(4)}$(at vanishing wavenumber)
\begin{eqnarray}
\tilde{\Gamma}_R^{(2)} ({\bf{k}}, \xi, u, \mu, \tilde{a}, d) & \; = \; & 
Z_\varphi \; \tilde{\Gamma}^{(2)}({\bf{k}}, \xi, \mu^{\epsilon} J^2_0
Z_u  \; u, \tilde{a}, d ) \quad ,\\[1cm]
\tilde{\Gamma}_R^{(4)} (\xi, u, \mu, \tilde{a}, d) & \; = \; & 
Z_\varphi^2 \; \tilde{\Gamma}^{(4)}( \xi, \mu^{\epsilon} J^2_0
Z_u \; u, \tilde{a}, d ) \quad .
\end{eqnarray}
The $Z$ factors $Z_\varphi$ and $Z_u$
can be determined for $d \leq 4$ by standard renormalization conditions at 
$\xi = \mu^{-1}$
\begin{equation}
\frac{\partial}{\partial k^2} \tilde{\Gamma}^{(2)}_R ({\bf{k}}, \mu^{-1}, u,
\mu, \tilde{a}, d)  _{\Big | {\bf{k}} = {\bf{0}}} \; = J_0 \quad ,
\end{equation}
\begin{equation}
\tilde{\Gamma}^{(4)}_R (\mu^{-1}, u, \mu, \tilde{a}, d) \; = \; 24 \; 
J^2_0 \; 
\mu^{\epsilon} u \quad .
\end{equation}
The $Z$ factors are finite for $d \leq 4$ if $\tilde{a} > 0$ and remain finite
for $d < 4$ if $\tilde{a} \to 0$ at fixed $u$ and $\mu$.
The following analysis is valid for $d < 4$ and $d = 4$  since
we keep the lattice spacing $\tilde{a}$ finite. 
Substituting
(20), (21), (24), (29) and (30) into (31) and (32) yields the Z factors in
one-loop order
\begin{eqnarray}
Z_\varphi (u, \mu \tilde{a}, d) &\; = \;& 1 \; + \; O  (u^2)\; ,\\[1cm]
Z_u  (u, \mu \tilde{a}, d) &\; = \;& 1 \; + \; 4(n+8)\; u \; I
(\tilde{a} \mu, d) \; + \;  O (u^2)
\end{eqnarray}
where
\begin{equation}
I(\tilde{a} \mu, d) \; = \; \mu^{4-d} \int\limits_{\bf{k}}
\left[ \mu^2 \; + \; \hat{J}_{\bf{k}}/J_0 \right]^{-2} \quad .
\end{equation}
We derive renormalization-group equations (RGE) for $\tilde{\Gamma}^{(N)}_R$
by taking the derivative of (29) and (30) with respect to 
$\mu$ at fixed 
$u_0, \tilde{a}$ and $r_0-r_{0c}$, i.e., at fixed $\xi$. 
This yields (at ${\bf{k}} = {\bf{0}}$)
\begin{equation}
\left[\mu \partial_\mu \; + \; \beta_u \partial_u +
\frac{N}{2} \zeta_\varphi   \right]
\tilde{\Gamma}^{(N)}_R (\xi, u, \mu, \tilde{a}, d) = 0
\end{equation}
with 
\begin{eqnarray}
\zeta_\varphi(u, \mu \tilde{a}, d) &\; = \;& 
(\mu\;  \partial_\mu \; \ln Z^{-1}_\varphi )_0 \quad ,\\[1cm]
\beta_u (u, \mu \tilde{a}, d) &\; = \;& 
(\mu\;  \partial_\mu \;  u)_0 \quad ,
\end{eqnarray}
where the index 0  
means differentiation at fixed parameters of the
bare theory. The formal solution of (36) reads 
\begin{equation}
\tilde{\Gamma}^{(N)}_R (\xi, u, \mu, \tilde{a}, d) \; = \; 
\tilde{\Gamma}^{(N)}_R (\xi, u (\ell), \ell  \mu, \tilde{a}, 
d) \exp \frac{N}{2}
\int\limits_1^\ell \; \zeta_\varphi(\ell^\prime) 
\frac{d\ell^\prime}{\ell^\prime}
\end{equation}
where $\zeta_\varphi(\ell) \equiv \zeta_\varphi (u(\ell), 
\ell \mu \tilde{a}, d)$ and where $u(\ell)$ is the solution of the 
flow equation
\begin{equation}
\ell \frac{d u(\ell)}{d \ell} \; = \; \beta_u (u(\ell), \ell \mu \tilde{a}, d)
\end{equation}
with $u(1) = u$. In the present context the most convenient choice of the
flow parameter $\ell$ is 
\begin{equation}
\ell \mu  \; = \; \xi^{-1} \quad .
\end{equation} 
We rewrite the renormalized vertex functions as 
\begin{equation}
\tilde{\Gamma}^{(N)}_R \; (\xi, u, \mu, \tilde{a}, d) \; = \; \xi^{-\delta_N}
f^{(N)} (\mu \xi, u, \mu \tilde{a}, d)
\end{equation}
where the amplitude functions $f^{(N)}$ are dimensionless and
\begin{equation}
\delta_N \; = \; d - (d-2) N/2 \; .
\end{equation}
The renormalizability of the $\varphi^4$ lattice model for $d \leq 4$
guarantees that the limit $\tilde{a} \to 0$ of $\tilde{\Gamma}^{(N)}_R$
at fixed $u, \mu$ and $\xi$ exists, i.e., that the function 
$f^{(N)}(\mu \xi, u, 0, d)$ is finite for
finite $\mu \xi$ and $u > 0$ for $d \leq 4$. From (39), (41) and (42) 
we obtain
\begin{equation}
f^{(N)} (\mu \xi, u, \mu \tilde{a}, d) \; = \; 
f^{(N)}(1, u(\ell), \tilde{a}/\xi, d) \exp \frac{N}{2} 
\int\limits_1^\ell \zeta_\varphi (\ell^\prime) 
\frac{d \ell^\prime}{\ell^\prime} 
\end{equation}
which provides the mapping of the amplitude
function $f^{(N)}(y,u, \mu \tilde{a}, d)$ from the critical region 
$y \gg 1$ (where perturbation theory breaks down)
to the noncritical value $y = 1$ (where perturbation theory is applicable). 
The temperature dependence of the amplitude function 
$f^{(N)} (1, u(\ell), \tilde{a}/\xi, d)$ in (44) is affected by the 
finite lattice constant $\tilde{a}$ not only through the explicit dependence
on $\tilde{a}/\xi$ but also through the effective coupling $u(\ell)$ that is
determined by the $\tilde{a}$ dependent RG flow equation (40). 
Furthermore the lattice constant enters the form of the temperature 
dependence of $\xi(t)$ according to (A.11) of Appendix A.

\subsection{Asymptotic Behavior}

Asymptotically $(\ell \to 0, \xi \to \infty)$
the effective coupling $u(\ell)$ approaches the fixed point $u^* = u(0)$
as determined by 
\begin{equation}
0 = \beta_u (u^*, 0, d)
\end{equation}
which is independent of the lattice constant $\tilde{a}$ and of the initial
value $u$. 
For $\xi \to \infty$, equation (44) approaches the asymptotic form
\begin{equation}
f^{(N)}(\mu \xi, u, \mu  \tilde{a}, d) \;  \sim \; 
A^{(N)} \; f^{(N)} (1, u^*, 0, d) (\mu \xi)^{N \eta/2}
\end{equation}
with the critical exponent
\begin{equation}
\eta = - \zeta_\varphi (u^*, 0, d)
\end{equation}
and the nonuniversal amplitude (which depends on $u$ and $\mu \tilde{a}$)
\begin{equation}
A^{(N)} = \exp 
\left\{ \frac{N}{2} \int\limits^0_1 \left[ \zeta_\varphi (\ell^\prime)\; - \;
\zeta_\varphi (0) \right] \frac{d \ell^\prime}{\ell^\prime}  \right\} \quad .
\end{equation}
The amplitude function $f^{(N)} (1,u,0,d)$ is finite and nonsingular at
$u = u^* > 0$ for $d < 4$
(compare the above statement after (43) regarding the renormalizability of 
$\tilde{\Gamma}^{(N)}_R$). We see that the dependence on the lattice
constant $\tilde{a}$ has disappeared asymptotically $(\tilde{a}/\xi \to 0)$
in the amplitude function on the right-hand side of (46).
Thus $\tilde{a}$ enters the asymptotic bulk critical behavior of 
$\tilde{\Gamma}^{(N)}_R$ only via $A^{(N)}$ (beyond one-loop order) which
is independent of $\xi$, and via the amplitude $\xi_0$ of $\xi$ [see (A.17)]. 
Therefore, taking the limit $\tilde{a} \to 0$ in the
renormalized quantity $\tilde{\Gamma}^{(N)}_R$ (as is usually done) is indeed
justified in the asymptotic bulk theory since this limit does not change 
the asymptotic
{\it{temperature}} dependence. For the asymptotic {\it{size}} dependence
of the  confined system, however, a corresponding property
is not generally valid, as we shall see in the subsequent Sections.

Application of the results to $N = 2$ yields, according to (8), the 
bare (physical) bulk susceptibility at ${\bf k} = {\bf 0}$
\begin{equation}
\chi_b \; = \; Z_\varphi (u, \tilde{a}/\xi_0, d)\; \xi^2 
\left[ f^{(2)} (1, u (\ell), \tilde{a}/\xi, d) \right]^{-1} \; \exp \;
\int\limits_\ell^1 \; \zeta_\varphi (\ell^\prime) 
\frac{d \ell^\prime}{\ell^\prime} \; .
\end{equation} 
The asymptotic $(\ell \to 0, \xi \to \infty )$ critical behavior 
above $T_c$ is
\begin{equation}
\chi_b \; = \; \tilde{A}_\chi \; \xi^{2-\eta} \; = \; A^+_\chi \; t^{-\gamma}
\quad .
\end{equation}
The amplitudes depend on $\tilde{a}$ according to 
\begin{eqnarray}
\tilde{A}_\chi & \; = \; & Z_\varphi (u, \tilde{a}/\xi_0, d) \; 
\xi^\eta_0 \; \left[ A^{(2)} \; f^{(2)}\; (1, u^*, 0,d) \right]^{-1} \quad ,
\\[1cm]
A^+_\chi &\; = \; & \xi^{2-\eta}_0 \; \tilde{A}_\chi \quad ,
\end{eqnarray}
where we have used  $\mu \; = \; \xi^{-1}_0$ and the asymptotic form
\begin{equation}
\xi \; = \; \xi_0 \; t^{-\gamma}
\end{equation}
with $\gamma = \nu (2-\eta)$. The dependence of $\xi_0$ on $\tilde{a}$
is given in (A.17)
of Appendix A.

\subsection{ Continuum Approximation}

For comparison we shall also consider the more standard version of the
$\varphi^4$ theory that is based on the continuum Hamiltonian \cite{14,Amit,13}
\begin{equation}
H \;  = \; \int\limits_V d^d x 
\left[ \frac{1}{2} r_0 \varphi^2  \; + \; \frac{1}{2} (\nabla \varphi)^2 
+ u_0 (\varphi^2)^2   \right]
\end{equation}
for the $n$-component field $\varphi({\bf{x}})$. Here the $(\nabla \varphi)^2$
term approximates the $(\varphi_i - \varphi_j)^2$ term of the lattice
Hamiltonian (7). The fluctuations of $\varphi({\bf{x}})$ are confined to
wavenumbers less than a finite cutoff $\Lambda$ corresponding to 
$\pi/\tilde{a}$. The bulk susceptibility corresponding to (9) is now 
defined by 
\begin{equation}
\chi_b ({\bf{k}}) \; = \; \lim_{L \to \infty} \quad \int\limits_V
d^d x \; < \varphi ({\bf{x}}) \; \varphi (0) > \; 
e^{-i{\bf{k}}\cdot {\bf{x}}} \quad .
\end{equation}
In the bulk limit, most expressions of the $\varphi^4$ lattice theory 
at finite $\tilde{a}$ 
remain applicable also to
the $\varphi^4$ field theory at finite $\Lambda$ after the replacements
$\hat{J}_{\bf{k}} \to {\bf{k}}^2$, $J_0 \to 1$ and $\tilde{a} \to \pi/\Lambda$ 
have been made. 
This implies that the asymptotic critical {\it temperature} dependence of
the $\varphi^4$ field and lattice theory for bulk systems is identical
(apart from different nonuniversal amplitudes). For confined systems, however,
a corresponding statement regarding the {\it size} dependence is not generally 
valid as we shall see in the  subsequent Sections.

\section{Renormalization Group and Finite-Size Scaling}

We are now in the position to discuss the size dependence of physical 
quantities within a renormalization-group treatment of the lattice model
(7) with a finite volume $V = L^d$. 
(For different geometries see below.)
We focus our analysis on the example
of the susceptibility above $T_c$
\begin{eqnarray}
\chi &\; = \;& \frac{\tilde{a}^{2d}}{L^d} \; \sum\limits_{i,j} \; 
<  \varphi_i  \; \varphi_j  > \\[1cm] 
&\; = \;& \chi (\xi, u_0, L, \tilde{a}, d) \quad .
\end{eqnarray}
Here we consider $\chi$ as a function of the bulk correlation length
$\xi$ (rather than of $r_0$) as explained in Section 2.

The basic assumption in the following is that, for periodic boundary 
conditions, 
the ultraviolet divergences of $\chi$ in the limit $\tilde{a} \to 0$
for the confined system are the same as those of the bulk susceptibility
$\chi_b$. This plausible assumption is in accord with the argument 
of Br\'{e}zin
\cite{6} regarding the decomposition of Fourier sums into bulk integrals
(which carry the ultraviolet divergent part) and finite-size contributions
(which are finite in the limit $\tilde{a} \to 0$). It is also in 
accord with one-loop results (Section 4) and with exact results in the 
large-$n$ limit (Section 5). Although this assumption has 
far-reaching consequences regarding the validity of finite-size scaling
for fixed finite $L/\xi$ we shall show that it does not rule out the 
possibility of a violation
of finite-size scaling
in the limit $L \gg \xi$.

Our assumption implies that the renormalized susceptibility $\chi_R$ 
at finite $L$ and at finite $\tilde{a}$ can be introduced for $d \leq 4$ as
\begin{equation}
\chi_R (\xi, u, L, \mu, \tilde{a}, d) \; = \; Z^{-1}_\varphi \;
\chi (\xi, \mu^{\epsilon} \; J^2_0\; Z_u \; u, L, \tilde{a}, d)
\end{equation}
where $Z_\varphi (u, \mu \tilde{a}, d)$ and $Z_u (u, \mu \tilde{a}, d)$ are
the {\it{bulk}} $Z$ factors defined in Section 2 and that $\chi_R$ remains finite
in the limit $\tilde{a} \to 0$ at fixed $\xi, u, L$ and $\mu$ for $d \leq 4$.
We note that there exists no justification {\it{a priori}}, however, 
to actually
perform this limit $\tilde{a} \to 0$ in the final results of the $\varphi^4$
theory if they are to be compared with those of model systems on lattices
with a finite lattice constant, e.g., with Monte Carlo data for Ising models.
Therefore we shall keep $\tilde{a}$ finite in the following analysis of the
size dependence of $\chi_R$ and of $\chi$.

We derive a renormalization-group equation for $\chi_R$ 
by taking the derivative of (58) with respect
to $\mu$ at fixed $u_0, \tilde{a}, L$ and $r_0 - r_{0c}$, i.e., at fixed
$\xi$. Since $L$ is not renormalized \cite{6} this yields 
\begin{equation}
\bigg[ \mu \partial_\mu \; + \; \beta_u \; \partial_u \; - \; \zeta_\varphi
\bigg] \chi_R \; (\xi, u, L, \mu, \tilde{a}, d) \; = 0 \; 
\end{equation}
where $\beta_u (u, \mu \tilde{a}, d)$ and $\zeta_\varphi (u, \mu \tilde{a}, d)$
are defined by (37) and (38) of the bulk theory. 
The solution reads for $d \leq 4$
\begin{equation}
\chi_R (\xi, u, L, \mu, \tilde{a}, d) \; = 
\; \chi_R (\xi, u(\ell), L, \ell \mu, \tilde{a}, d)\; \exp 
\int\limits^1_\ell \; \zeta_\varphi (\ell^{\prime}) \; 
\frac{dl^{\prime}}{\ell^{\prime}}
\end{equation}
where $\ell$ can be chosen arbitrarily in an exact theory. For the
purpose of an application to perturbative results a natural
choice is
\begin{equation}
\xi^{-2} \; + \; L^{-2} \; = \; \mu^2 \ell^2 \quad .
\end{equation}
Although the RGE (59) has the same form as the bulk RGE (36) the 
simultaneous
appearance of the {\it three} lengths $\xi, L$ and $\tilde{a}$ in the arguments
of $\chi_R$ in (59) and (60) complicates the situation and requires a careful
consideration of different limiting cases.
To answer the question about the possible relevance of the dependence on 
$\tilde{a}$ we distinguish the following three cases (i) - (iii).

(i) At $T = T_c$ or $\xi = \infty$ and  at finite $L$, we introduce
a dimensionless function $\tilde{f}_\chi$ according to
\begin{equation}
\chi_R (\infty, u, L, \mu, \tilde{a}, d) \; = \; L^2 \; \tilde{f}_\chi
(\infty, u, \mu L, \tilde{a}/L, d)\; .
\end{equation}
In the large-$L$ limit we obtain from (60) - (62) with 
$\ell = \mu^{-1} L^{-1}$ 
\begin{equation}
\chi_R (\infty, u, L, \mu, \tilde{a}, d) \; \sim \; L^2 (\mu L)^{-\eta}
\left[A^{(2)} \right]^{-1} \; \tilde{f}_\chi (\infty, u^*, 1, \tilde{a}/L, d)
\end{equation}
where $\eta$ and $A^{(2)}$ are given in (47) and (48). The remaining
dependence of $\tilde{f}_\chi$ on $\tilde{a}/L$ in (63) yields only a
subleading correction to the leading power law $\sim \; L^{2-\eta}$
provided that the function $f_\chi (\infty, u^*, 1, 0, d)$ is finite. The
latter property is valid for $d < 4$ (where $u^* > 0$) provided that 
 the $\varphi^4$ theory is renormalizable at finite $L$ and $\tilde{f}_\chi $ 
is non-singular at $T_c$ for finite $\mu L$,
i.e., provided that the limit (at fixed $u$ and $L$)
\begin{equation}
\lim_{\tilde{a} \to 0} \tilde{f}_\chi (\infty, u, 1, \tilde{a}/L, d)
\; = \; \tilde{f}_\chi (\infty, u, 1, 0, d)
\end{equation}
exists and is finite for $u > 0$ at $d \leq 4$.

(ii) For $0 < \xi < \infty$ at finite $L$, i.e., at finite ratio 
$0 < \xi/L < \infty$, we introduce the dimensionless amplitude 
function $f_\chi$ according to
\begin{equation}
\chi_R (\xi, u, L, \mu, \tilde{a}, d) \; = \; \xi^2 \; f_\chi (\mu \xi, u, 
\mu L,  \tilde{a}/L, d) \quad.
\end{equation}
In the asymptotic region $\xi \gg \tilde{a}, L \gg \tilde{a}$
corresponding to $\ell \ll 1$
we obtain from (60)
\begin{equation}
\chi_R (\xi, u, L, \mu, \tilde{a}, d) \; \sim \; \xi^2 \ell^\eta 
\left[ A^{(2)} \right]^{-1} \;  f_\chi (\ell \mu \xi, u^*, 
\ell \mu L,
\tilde{a}/L, d)\quad .
\end{equation}

Renormalizability of the $\varphi^4$ theory at finite $L$ guarantees 
that the limit
\begin{equation}
\lim_{\tilde{a} \to 0} f_\chi (\mu \xi, u, \mu L, \tilde{a}/L, d) \; = \;
f_\chi (\mu \xi, u, \mu L, 0, d)
\end{equation}
exists and is finite for finite arguments and $d \leq 4$.
Therefore, taking the limit $L \to \infty$ in (66) at fixed finite ratio
$0 < L/\xi < \infty$, i.e., at fixed  values of 
$\ell \mu \xi$ and $\ell \mu L$, yields a finite amplitude function for 
$d < 4 \; (u^* > 0)$
\begin{equation}
f_\chi (\ell \mu \xi, u^*, \ell \mu L, 0, d) \; = \; Y (L/\xi)
\end{equation}
on the right-hand side of (66). Here we have used the fact that 
$\ell \mu \xi$ and $\ell \mu L$ depend only on $L/\xi$
according to (61).
Thus the dependence on $\tilde{a}/L$ in (66) represents only a subleading
correction to the leading size-dependence (68) provided that $L/\xi$ is
finite. This implies that both $\chi \; = \; Z_\varphi (u, \mu \tilde{a}, d) 
\; \chi_R$ and $\chi_R$ attain the finite-size scaling form
\begin{equation}
\chi_R (\xi, u, L, \mu, \tilde{a}, d) \; \sim \; \xi^{2-\eta} \mu^{-\eta}
(1 + \xi^2/L^2)^{\eta/2} \left[ A^{(2)}   \right]^{-1} \; Y (L/\xi)
\end{equation}
in the asymptotic region $L \gg \tilde{a}$ and $\xi \gg \tilde{a}$ for
any finite ratio $0 < L/\xi < \infty$. This is in agreement with Br\'{e}zin's
conclusion \cite{6} who performed the limit $\tilde{a} \to 0$ at the outset.

(iii) There exist, however, significant paths in the asymptotic
$L^{-1} - \xi^{-1}$ plane (Fig. 1) along which the ratio $L/\xi$ does not
remain finite but diverges. These paths include the approach 
towards asymptotic bulk critical behavior at fixed
$\tilde{a}/\xi > 0$ (arrow in Fig. 1). This case is not covered
by the discussion of case (ii) above and was not considered in Br\'{e}zin's
analysis \cite{6}. In this regime $(L/\xi \gg 1, \tilde{a}/\xi > 0)$
we make   the choice
\begin{equation}
\ell \; = \; \mu^{-1} \xi^{-1}
\end{equation} 
instead of (61). From (60), (66) and (70) we then obtain asymptotically
\begin{equation}
\chi_R (\xi, u, L, \mu, \tilde{a}, d) \; \sim \; \xi^{2-\eta} \mu^{-\eta}
\left[ A^{(2)}  \right]^{-1} \;
f_\chi (1, u^*, L/\xi, \tilde{a}/L, d) \; .
\end{equation}
In the bulk limit $L \to \infty$ at fixed $\xi < \infty$, 
equations (58) and (71) agree with (50) and (51) where
\begin{equation}
f_\chi (1, u^*, \infty, 0,d) \; = \; \left[ f^{(2)} (1,u^*, 0, d) \right]^{-1}
\; \equiv \; f^*_b
\quad .
\end{equation}
Similar to the case (ii), the renormalizability of the $\varphi^4$ theory
at finite $L$ guarantees that in the limit $\tilde{a} \to 0$ the function
\begin{equation}
\lim_{\tilde{a} \to 0} f_\chi \; (1, u^*, L/\xi, \tilde{a}/L, d) \; 
= \; f_\chi (1,  u^*, L/\xi, 0, d) \; \equiv \; f^*_b \; - \; f_1 (L/\xi)
\end{equation}
exists and is finite for finite arguments (with $u^* > 0$ for 
$d < 4$ dimensions). 
Thus, at first sight, the dependence of $f_\chi$ on $\tilde{a}/L$ on the
right-hand side of (71) appears to be a subleading correction that can
be neglected in the asymptotic region $\tilde{a}/L \ll 1$, similar to the case
(ii). 
A closer inspection shows, however, that this reasoning
is not compelling in the present case (iii) where $L/\xi$ may become 
arbitrarily large at fixed finite $\xi$.

On a formal level this is seen by rewriting the dimensionless
function $f_\chi$ as
\begin{equation}
f_\chi (1, u^*, L/\xi, \tilde{a}/L, d) \; = \; F_\chi (L/\xi, \tilde{a}/\xi)
\end{equation}
where the dependence of $F_\chi$ on $\tilde{a}$ appears in the form
$\tilde{a}/\xi$ rather than $\tilde{a}/L$. Now there exists no argument
why the dependence on $\tilde{a}/\xi$ should be negligible in the
limit $L/\xi \gg 1$ at fixed $\tilde{a}/\xi > 0$. More specifically,
consider the decomposition
\begin{equation}
F_\chi (L/\xi, \tilde{a}/\xi) \; = \; f^*_b \; - \; f_1 (L/\xi)
\; - \; f_2 (L/\xi, \tilde{a}/\xi) \quad 
\end{equation}
where $f^*_b$ and $f_1$ are independent of $\tilde{a}$ as defined in (72)
and (73).
The last term $f_2$ contains the complete $\tilde{a}$ dependence 
of $f_\chi$ and
vanishes for $\tilde{a} \to 0$.

Clearly only the term $f_1(L/\xi)$ in (75) is in agreement with
finite-size scaling in contrast to the  term 
$f_2 (L/\xi, \tilde{a}/\xi)$.
Thus the fundamental question arises whether the size dependence of $f_2$
at finite
$\tilde{a}/\xi$ and for large $L/\xi$ 
is asymptotically negligible compared to that of $f_1$. 
Only if the ratio 
\begin{equation}
R(L/\xi, \tilde{a}/\xi)
\; = \; \frac{f_2(L/\xi, \tilde{a}/\xi)}{f_1 (L/\xi)} 
\end{equation}
of equations (3)-(6) would vanish as $L/\xi \to \infty$ at fixed 
$0 < \tilde{a}/\xi < \infty$ there would be no violation of finite-size 
scaling.
Renormalizability
guarantees the existence of the function $f_1(L/\xi)$ 
but does not say anything about
the magnitude and sign of the ratio $R(L/\xi, \tilde{a}/\xi)$ 
in the regime $L/\xi \gg 1$.
Although it is clear that the {\it{total}} finite-size deviation
from asymptotic bulk critical behavior [compare (2)]
\begin{eqnarray}
\Delta \chi   \equiv \frac{\chi_b - \chi}{\chi_b}  
&\; =  \; & \frac{ f^*_b \; - \; f_\chi (1, u^*, L/\xi, \tilde{a}/L,d) 
}{f^*_b}\\[1cm] 
&\; = \;& \frac{ f_1 (L/\xi)}{f_b^*} \bigg[ 1 \; + \; R (L/\xi, 
\tilde{a}/\xi) \bigg]
\end{eqnarray}
must approch zero as $L/\xi$ increases, renormalizability does not rule
out the possibility that the {\it{relative}}
contribution described by $|R|$ increases and becomes large compared to $1$ 
with increasing 
$L$ at fixed
$\xi < \infty$ and $\tilde{a} > 0$ in the region $L \gg \xi$.
In fact it does not even rule out the possibility that 
$R(L/\xi, \tilde{a}/\xi)$
{\it{diverges}} as $L/\xi \to \infty$ at fixed $0 < \tilde{a}/\xi < \infty$.
If this is the case then $f_2$ becomes dominant compared to $f_1$ in (75) and 
finite-size scaling is violated in the lower part of the $L^{-1} - \xi^{-1}$ 
plane close to the bulk limit (Fig. 1).

We illustrate these considerations by a simple example: if $f_2 (x,y) $ would
be $x y$ for small $y$ and for general (arbitrarily large) $x$, then one 
should not dismiss $f_2 (x,y)$
as a correction that is negligible (compared to $f_1 (x)$) for small $y > 0 $
since $f_2$ can become large for $x \gg 1/y$. Actually we shall 
specify $f_2 (x,y)$ essentially as an exponential function of $xy^2$ 
[see equations (103) and (104)].

No general arguments but only explicit calculations can answer our
question about the magnitude and sign of the {\it{nonuniversal}} quantity 
$R$. In Section 4 we shall calculate
$R$ for general $n$ in one-loop order. In Section 5 the exact form of 
$R$  will be derived in the large-$n$ limit.
We shall indeed show that $R$ diverges, 
\begin{equation}
\lim_{x \to \infty} R(x, \tilde{a}/\xi) \; = \; 
\infty, \; \quad  \tilde{a}/\xi \; \mbox{fixed} \quad \; , \;  
\tilde{a}/\xi > 0 
\end{equation}
at any finite $\tilde{a}/\xi$ for the $\varphi^4$ lattice
model below four dimensions. Corresponding properties remain valid
also in the cases $d = 4$ and $d > 4$ whose consequences 
will be studied elsewhere \cite{CDnew}.

The analysis of this Section can be repeated for the field-theoretic 
$\varphi^4$ model at finite cutoff $\Lambda$. The reasoning remains of
course parallel to that given above and leads to the question about the
magnitude and sign of the ratio 
$R_{field} (L/\xi, \Lambda \xi)$. We shall show in 
Sections 4 and 5 that 
$R_{field}$  differs fundamentally from 
$R$ 
and that in both cases finite-size scaling is violated for 
$L/\xi \gg 1$.

Finally we note that the analysis of this Section did not make explicit
use of periodic boundary conditions except that the renormalizability in 
terms of bulk $Z$ factors was assumed according to (58). Therefore our line
of thoughts should remain applicable more generally to those cases
where the bulk renormalizations suffice to renormalize the physical 
quantities of the confined system. Our analysis should  also be extended to 
the important case where additional (surface) renormalizations come into 
play which, for confined systems, have been studied so far only in the 
continuum approximation (54) ( with surface terms) and only in the 
limit $\Lambda \to \infty$.

\section{ Perturbation Theory above ${\bf{T_c}}$}

As is well known a calculation of finite-size effects within the
$\varphi^4$ theory including the size-dependence at $T_c$ requires a
decomposition into modes where the lowest mode is separated and only
the higher modes are treated perturbatively \cite{7,8}. 
As noted recently \cite{15}, however, this perturbative calculation does
not correctly capture the exponential size dependence of the approach
towards bulk critical behavior within the $\varphi^4$ lattice model.
Covering the size dependence in the entire $L^{-1} - \xi^{-1}$ plane,
i.e., both at $T = T_c$ as well as for $L/\xi \gg 1$ at fixed $\xi < \infty$,
would require a non-perturbative treatment \cite{15} within the mode expansion
mentioned above. Such a treatment could be given on the basis of the 
order-parameter distribution function \cite{12} that includes the
higher modes in a non-perturbative way.

Here we point out, however, that such a non-perturbative treatment can be 
avoided because the exponential size dependence of the $\varphi^4$ 
lattice model above $T_c$ is not of a truly non-perturbative nature, 
unlike the effects due to the Goldstone modes
below $T_c$ \cite{12} . We shall show that an ordinary renormalized 
perturbation approach is sufficient if one restricts oneself to the 
region $L > \xi$ above $T_c$. This is just the region where the nonuniversal
finite-size effects due to a finite lattice constant become significant.
In this region a separation between the lowest mode and the higher modes
is unnecessary since the lowest mode does not become
dangerous in the bulk limit $L \to \infty$ at fixed $\xi < \infty$. 
It does not even become dangerous in the limit $L \to \infty, \xi \to \infty$ 
at fixed finite ratio $L/\xi > 0$. Therefore we do not separate the lowest
mode but instead shall
present an ordinary perturbation approach, similar to bulk perturbation
theory, where all modes are treated in the same way. Although this 
approach deteriorates with increasing $\xi/L$ and breaks down in the 
region $\xi \gg L $ due to the dangerous lowest mode, it 
is well applicable
for $L > \xi$.

\subsection{ Lattice Model at Finite Lattice Constant}

We start out from the $\varphi^4$ lattice Hamiltonian (7) (for a
$d$ dimensional cube with $V = L^d$ and periodic boundary conditions) 
in the Fourier representation
\begin{eqnarray}
H \; & = & \;  L^{-d} \sum\limits_{\bf{k}} \frac{1}{2} 
\left[ r_0 \; + \; \hat{J}_{\bf{k}}  \right] \;
\varphi_{\bf{k}} \varphi_{\bf{-k}} \nonumber \\ [1cm]
& + & u_0 L^{-3d} \; {\sum\limits_{\bf{k k^\prime k^{''}}}}
\left( \varphi_{\bf{k}} \varphi_{\bf{k^\prime}} \right) 
\left( \varphi_{\bf{k^{''}}} \varphi_{\bf{\bf{-k-k^\prime-k^{''}}}} \right) 
\quad ,  \\[1cm]
\varphi_{\bf{k}} \;& = & \; \tilde{a}^d \; \sum\limits_j e^{-i{\bf{k \cdot x}}_j}
\varphi_j   
\end{eqnarray}
where $\hat{J}_{\bf{k}}$ is defined by (22) and (23) but now for the finite
lattice.
The summations in (80) run over discrete ${\bf{k}}$ vectors with
components $k_j  \; = \;  2 \pi m_j/L \; , \; m_j \; = \; 0, \; \pm \; 1, \; 
\pm \; 2, \; \cdots , \; j \; = \; 1, 2, \cdots , d $
in the range $- \pi/\tilde{a} \; \leq \; k_j \; < \; \pi/\tilde{a}$
with a finite lattice spacing $\tilde{a}$.

The standard one-loop expression for the (inverse) susceptibility (56) above
$T_c$
reads for the finite system
\begin{equation}
\chi^{-1} \; = \; r_0 \; + \; 4(n + 2) \; u_0 \; L^{-d} \; \sum\limits_{\bf{k}}
\; (r_0 \; + \; \hat{J}_{\bf{k}})^{-1} \; + \; O(u^2_0) 
\end{equation}
and
\begin{equation}
\chi^{-1}_b \; = \; r_0 \; + \; 4(n + 2) \; u_0 \; \int\limits_{\bf{k}}
\; (r_0 \; + \; \hat{J}_{\bf{k}})^{-1} \; + \; O(u^2_0) 
\end{equation}
for the bulk system. Since $\hat{J}_{\bf{k}}$ is a periodic 
function of each component $k_j$ the one-loop sum in (82) satisfies the Poisson
identity at finite $\tilde{a}$ \cite{6,Poisson}
\begin{equation}
L^{-d} \; \sum\limits_{{\bf{k}}} \; (r_0 \; + \; \hat{J}_{\bf{k}})^{-1}
\; = \; \sum\limits_{\bf{n}} \; \int\limits_{\bf{k}} \; (r_0 
\; + \; \hat{J}_{\bf{k}})^{-1} \; e^{i{\bf{k}} \cdot {\bf{n}} L}
\end{equation}
where ${\bf{k}} \cdot {\bf{n}} \; = \; \sum_j \; k_j \; n_j$ . 
The sum $\sum_{\bf{n}}$ runs over all integers $n_j, \; j = 1,2,\cdots,d$
in the range $- \infty  \leq n_j \leq \infty$ (whereas $\sum_{\bf{k}}$ and
$\int_{\bf{k}}$ have  finite cutoffs $\pm \pi/\tilde{a}$). 
The crucial quantity that contains
all finite-size effects is the function
\begin{eqnarray}
D(r_0, L, \tilde{a})  &\; = \;& 
L^{-d} \; \sum\limits_{\bf{k}} (r_0 
+ \hat{J}_{\bf{k}})^{-1} \quad - \quad 
\int_{\bf{k}} (r_0 \; + \; \hat{J}_{\bf{k}})^{-1} \\[1cm]
&\; = \; &  \sum\limits_{ {\bf{n}} \neq  {\bf{0}} } \; \int\limits_{\bf{k}}
(r_0 + \hat{J}_{\bf{k}})^{-1} \; e^{i{\bf{k}} \cdot {\bf{n}} L} \quad .
\end{eqnarray}
The sum $\sum\limits_{\bf{k}}$ in (85) includes the lowest-mode (${\bf{k}} = 
{\bf{0}}$) term
$L^{-d} \; r_0^{-1}$.
It is only this lowest-mode term that diverges for $r_0 \to 0$ at finite $L$
whereas the ${\bf{k}} \neq {\bf{0}}$ contributions remain finite 
in this limit. 
For finite $r_0 L^2 > 0, D(r_0, L, \tilde{a})$  is of $O(L^{2-d})$ whereas 
for $r_0 L^2 \gg 1$
it is of $O (\exp - L \;  r_0^{1/2})$ (see below).
A simple rearrangement yields 
\begin{equation}
\chi (\xi, u_0, L, \tilde{a}, d)^{-1} \; = \; J_0 \xi^{-2} \; \left[ 1 \; + 
\; 4(n + 2)u_0 \;J^{-2}_0 \;  
\xi^2  \tilde{D}  \quad + \quad O (u_0^2) \right]
\end{equation}
with $\tilde{D} \; = \; J_0 \;  D (J_0 \xi^{-2}, L, \tilde{a})$, i.e.,
\begin{equation}
\tilde{D} (\xi, L, \tilde{a}) \; = \;  
L^{-d} \; 
\sum\limits_{\bf{k}}
\;(\xi^{-2} \; + \; \hat{J}_{\bf{k}}/J_0)^{-1}
\quad - \quad
\int\limits_{\bf{k}} \; (\xi^{-2} \; + \; \hat{J}_{\bf{k}}/J_0)^{-1}
\end{equation}
where we have used $\chi_b^{-1} \; = \; J_0 \; \xi^{-2} \; + \; O (u^2_0)$
on the level of bare perturbation theory according to (8) and (20).

The quantity $\tilde{D}$ 
remains finite in the limit $\tilde{a} \to 0$ for finite $\xi$ in  
arbitrary dimensions. 
This means that the ultraviolet divergence of $\chi$ at finite $L$ in
one-loop order is absorbed by the {\it{bulk}} correlation length $\xi$, 
in accord with the assumption in Section 3.
For $\tilde{a} \to 0$  and $L/\xi > 0$, the function $\tilde{D}$ 
can be represented as  \cite{7,8}
\begin{equation}
\tilde{D}(\xi, L, 0) \; = \; L^{2-d} \; I (L^2/\xi^2)
\end{equation}
where
\begin{equation}
I(x) \; = \; (4 \pi^2)^{-1} \; \int\limits^{\infty}_0 \; dz \; e^{-xz/4\pi^2}
\left[  K(z)^d  \; - \; \left( \frac{\pi}{z}    \right)^{d/2}  \right]
\end{equation}
with
\begin{equation}
K(z) \; = \; \sum\limits_{m = - \infty}^{\infty} \; \exp
(-zm^2) \quad .
\end{equation}
$I(x)$ diverges for $x \to 0$ due to the lowest-mode term but is
exponentially small for $x \gg 1$.
For $\tilde{a} > 0$ we decompose $\tilde{D}$  as
\begin{equation}
\tilde{D} (\xi, L, a) \; = \; L^{2-d} f_D (L/\xi, \tilde{a}/\xi) \; ,
\end{equation}
\begin{equation}
f_D (L/\xi, \tilde{a}/\xi) \; = \; 
I(L^2/\xi^2) \; + \; M (L/\xi \; , \; 
\tilde{a}/\xi)  \; ,
\end{equation}
where  $M (L/\xi, \tilde{a}/\xi) $ contains the $\tilde{a}$ dependence of
$\tilde{D}$ and vanishes for $\tilde{a} \rightarrow 0$. The explicit form of 
$M (L/\xi, \tilde{a}/\xi) $ will be determined by equations (98), (101), (103)
and (104) below.

We note that $\tilde{D}$ does not require a renormalization as it depends
only on $L, \xi$ and $\tilde{a}$. 
Application of the RG analysis of the preceding Section to (87) yields
\begin{equation}
\chi \; = \; Z_\varphi (u, \mu \tilde{a}, d) \;
f_\chi (\ell \mu \xi, u (\ell), \ell \mu L, \tilde{a}/L,d) \; \xi^2
\exp \int\limits^1_\ell \zeta_\varphi (\ell^\prime) 
\frac{d \ell^\prime}{\ell^{\prime}}
\end{equation}
where in one-loop order
\begin{equation}
f_\chi (\mu \xi, u, \mu L, \tilde{a}/L, d) \; = \; J^{-1}_0 
\left\{ 1 - 4 (n + 2) u \mu^{\epsilon} \xi^{\epsilon} 
(\xi/L)^{d-2} f_D (L/\xi, \tilde{a}/\xi) \; + \; O(u^2) 
\right\}.
\end{equation}
For the application to $L/\xi \gg 1$ at $\tilde{a}/\xi > 0$ we choose the
flow parameter as $\ell = \mu^{-1}\xi^{-1}$. Then the finite-size deviation
$\Delta \chi$  from the bulk susceptibility
\begin{equation}
\chi_b \; = \; Z_\varphi (u, \mu \tilde{a}, d) 
\bigg\{J_0^{-1} \; + \; O \left[ u(\ell)^{2} \right]
\bigg\} \; \xi^2 \exp \int\limits^1_\ell \; \zeta_\varphi (\ell^\prime)
\frac{d \ell^\prime}{\ell^\prime}
\end{equation}
becomes in one-loop order
\begin{equation}
\frac{\chi_b - \chi}{\chi_b} \; = \; 4(n+2)u(\ell) (\xi/L)^{d-2}
\; I(L^2/\xi^2)
\Big[ 1 \; + \; R(L/\xi, \tilde{a}/\xi) \Big] \; + \; O 
\left[ u(\ell)^2 \right]
\end{equation}
with
\begin{equation}
R(L/\xi, \tilde{a}/\xi) \; = \; \frac{M(L/\xi, 
\tilde{a}/\xi)}{I(L^2/\xi^2)}\; .
\end{equation}
These results are valid at finite $\tilde{a}$ for both $d < 4$ and $d = 4$
dimensions and still contain all non-asymptotic contributions 
(Wegner corrections \cite{Wegner}) within
the $\varphi^4$ model. These contributions enter through $u(\ell)$ as well
as through the non-asymptotic form of $\xi$ as a function of $t$ as determined
by (A.11) of Appendix A. Sufficiently close to $T_c$ such non-asymptotic 
contributions become negligible. By contrast, the nonuniversal term $R(L/\xi,
\tilde{a}/\xi)$ {\it{cannot be considered as a non-asymptotic contribution}}
since it is nonnegligible at any $T > T_c$ for sufficiently large $L$ as we
shall see below.

Neglecting the deviation of $u(\ell)$ from the fixed point value $u^*$
we obtain for $d < 4$ 
\begin{equation}
\Delta \chi \; \equiv \; \frac{\chi_b - \chi}{\chi_b} \;  
= \; g(L/\xi) \; \Big[ 1 \; + \; R(L/\xi, \tilde{a}/\xi)\Big]
\end{equation}
with the universal part
\begin{equation}
g(L/\xi) \; = \; 4(n + 2) \; u^* \; (\xi/L)^{d-2}  
I(L^2/\xi^2) \; + \; O (u^{*2}) \; .
\end{equation}
For $L/\xi \gg 1$ the function $I(L^2/\xi^2)$ becomes
\begin{equation}
I(L^2/\xi^2) \; = \; d \; (2\pi)^{(1-d)/2} (L/\xi)^{(d-3)/2} \;  \exp(-L/\xi)
\quad .
\end{equation}

The nonuniversal part $R(L/\xi, \tilde{a}/\xi)$ 
depends on the detailed form of 
$\hat{J}_{\bf{k}}$.
For simplicity we assume a simple-cubic lattice
with nearest-neighbor coupling $J$,
\begin{equation}
\hat{J}_{\bf{k}} \; = \; \frac{4 J}{\tilde{a}^2} \; \sum\limits_{j = 1}^{d}
\; \left[ 1 \; - \cos (\tilde{a} k_j)  \right]  \quad , 
\end{equation}
which implies $J_0 \; = \; 2 J$. 
A calculation parallel to that in Appendix A of Ref. \cite{21}  
yields for $L \gg \xi  \gg 
\tilde{a}$
\begin{equation}
R(L/\xi, \tilde{a}/\xi) \; = \; \exp \left[ \Gamma (\tilde{a}/\xi)
\; \frac{L}{\xi} \right] \; -1 
\end{equation}
with the function
\begin{equation}
\Gamma (\tilde{a}/\xi) \; = \; \frac{1}{24} (\tilde{a}/\xi)^2 \; + \; 
O \left[ (\tilde{a}/\xi)^3 \right] \; .
\end{equation}
We see that, at any fixed $\tilde{a}/\xi > 0, \; R$
diverges for $L/\xi \to \infty$. Thus the term $R(L/\xi, \tilde{a}/\xi)$ 
is nonnegligible for sufficiently large $L$, even arbitrarily 
close to $T_c$, unlike the Wegner corrections \cite{Wegner}
arising from the  
deviation of the effective coupling $u(\ell)$ in (97)  from its fixed
point value $u^*$. The resulting asymptotic size
dependence of $\Delta \chi$ for large $L/\xi$ and fixed $\tilde{a}/\xi > 0$
is determined by 
\begin{equation}
\frac{\chi_b - \chi}{\chi_b} \; \sim  
\; g(L/\xi) \; R \; (L/\xi, \tilde{a}/\xi) 
\qquad \qquad \qquad  \qquad \qquad
\end{equation}

\begin{equation} 
=   \;
4 (n + 2) \; u^* \; d \;  (2 \pi)^{(1-d)/2} 
 \;  (L/\xi)^{(1-d)/2} \; \exp \; 
\left[ \frac{1}{24}(\tilde{a}/\xi)^2  \; \frac{L}{\xi}  \right] \exp (-L/\xi)
\end{equation}
with $u^* > 0$ for $d < 4$.
This behavior
is nonuniversal and depends on the {\it{two}} 
ratios $L/\xi$ and $\tilde{a}/\xi$
in an essential way, rather than only on $L/\xi$. 
A dominant influence of $\tilde{a}/\xi$ exists in the non-scaling region where
$R > 1$ corresponding to the region below the dashed line in
the $L^{-1} - \xi^{-1}$ plane in  Fig. 1.
This line is determined  by $R = 1$, i.e., 
\begin{equation}
\tilde{a}/L \; = \; [24 \ln 2]^{-1} \; (\tilde{a}/\xi)^3 \quad .
\end{equation}

Non-negligible effects arising from $R$ exist already above this line.

The exponential part of (106) could be written in the form  
$\exp(-L/\xi_{eff})$ with $\xi_{eff}=\xi/[1- (\tilde{a}/\xi)^2/24 + ...]$ 
which then would hide the 
violation of finite-size scaling [whereas the violation is quite explicit in 
the form of (103)-(106)]. But the true bulk correlation length $\xi$ 
(including all non-asymptotic bulk corrections) is already precisely 
defined by (10) and cannot be
arbitrarily redefined here to comply with scaling.

As a remarkable feature we note that, in one-loop order, the
nonuniversal  function
$R$ and the condition (107) are independent of the dimension $d$, of the
number of components $n$ and of the fixed point value $u^*$ of the
four-point coupling. Therefore the same function $R(L/\xi, \tilde{a}/\xi)$
causes a violation of the two-variable finite-size scaling form 
[59,60,63,66-68] above four dimensions as will be further discussed
elsewhere \cite{CDnew}.

These results can be easily generalized to a $d$ dimensional system 
with a partially finite
geometry that is confined in $\tilde{d}$ dimensions (size $L$) and is
infinite in $d - \tilde{d}$ dimensions. This includes the cubic, film
and cylindrical geometries as special cases $\tilde{d} = d, \tilde{d} = 1$
and $\tilde{d} = d - 1$, respectively. 
Instead of (82) we then have
\begin{equation}
\chi^{-1} \; = \; r_0 + 4(n+2)u_0 \; L^{-\tilde{d}}  
\sum\limits_{\bf{q}} \;
\int\limits_{\bf{p}} (r_0 + \hat{J}_{\bf{k}})^{-1} \; + \;
O(u^2_0)
\end{equation}
where the $d$-dimensional vector ${\bf{k}} = ({\bf{q}}, {\bf{p}})$ has
$\tilde{d}$ components ${\bf{q}} = (q_1, ..., q_{\tilde{d}})$ 
and $d - \tilde{d}$
components ${\bf{p}} = (p_{\tilde{d} + 1)}, ... , p_d)$. Equation (88) is
modified accordingly. The integral representation of
$I(L^2/\xi^2)$, (90), is replaced by
\begin{equation}
I(x) \; = \; (4 \pi^2)^{-1}
\; \int\limits^{\infty}_0 dz \; e^{-xz/4\pi^2}\;
\left[ K(z)^{\tilde{d}}  \;
\left( \frac{\pi}{z} 
\right)^{(d-\tilde{d})/2} \; 
- 
\left( \frac{\pi}{z} 
\right)^{d/2} \right] .
\end{equation}
For $L/\xi \gg 1$ this yields, instead of (101), 
\begin{equation}
I(L^2/\xi) \; = \; \tilde{d} \;  (2 \pi)^{(1-d)/2}\; (L/\xi)^{(d-3)/2} \;
\exp (-L/\xi) \;.
\end{equation}
The function $R(L/\xi, \tilde{a}/\xi)$, however, remains unchanged, i.e., it
is independent of the geometry (in one-loop order). 
Thus the leading finite-size deviation $\Delta \chi$ for large
$L/\xi$ and fixed $\tilde{a}/\xi > 0$ for general $\tilde{d} \leq d < 4$ is 
given by
\begin{equation}
\frac{\chi_b - \chi}{\chi_b} \; \sim \; 4(n + 2) u^* \tilde{d}\; 
(2 \pi)^{(1-d)/2}
(L/\xi)^{(1-d)/2} \; \exp \left[ \frac{1}{24} (\tilde{a}/\xi)^2 \frac{L}{\xi}
 \right] \exp (-L/\xi) \; .
\end{equation}
This differs from (106) only by the replacement $d \to \tilde{d}$
in the prefactor.

We conclude that  finite-size scaling is
violated below four dimensions in the $\varphi^4$ lattice model with
periodic boundary conditions  above $T_c$
in one-loop order for general $n$ in the region $L/\xi \gg 1$ at any
finite $\xi < \infty$ even arbitrarily close to $T_c$. 
Clearly higher-loop contributions cannot
remedy this violation.

Although one should trust the perturbative results of the $\varphi^4$
theory primarily  for $2 < d \leq 4$ one cannot exclude the possibility
that an extrapolation to $d = 2$ yields sensible results, possibly for
general $n$ above $T_c$ and for $n = 1$ below $T_c$. This appears to
be suggestive for the structure of our results (97), (106) and (111). 
This could be of particular relevance
for the case $n = 1$ for which exact results of the two-dimensional 
Ising model in a confined geometry are available \cite{16,AY}. In all cases,
contributions with  an exponential size dependence were found above $T_c$
for $L \gg \xi$. It would be interesting to reanalyze these results 
\cite{16,AY}
including all exponential and non-exponential prefactors which so far have 
not been worked out explicitly (see, e.g., equation (6.8) of Ref. \cite{2})
and to compare the structure of these results with that of our equations (106)
and (111). Our results suggest that these prefactors may contain  
{\it{nonuniversal}} contributions (such as our $\tilde{a}$ dependent 
exponential factor)  that cannot be neglected in the limit 
$L/\xi \gg 1$ at fixed $\tilde{a}/\xi > 0$.

\subsection{ Field-Theoretic Model at Finite Cutoff}

The susceptibility above $T_c$ of the field-theoretic model (54) is defined
as 
\begin{equation}
\chi \; = \; \int\limits_V \; d^dx \; 
< \varphi({\bf{x}}) \varphi ({\bf{0}}) > \; .
\end{equation}
The one-loop expression of the susceptibility for a cubic geometry, $V = L^d$,
reads [compare (87) and (88)]
\begin{equation}
\chi (\xi, u_0, L, \Lambda, d)^{-1} \; = \; \xi^{-2}
\left[ 1 + 4 (n + 2) u_0 \xi^2 \tilde{D}_{field} + O (u^2_0) \right]
\end{equation} 
where now
\begin{equation}
\tilde{D}_{field} (\xi, L, \Lambda) \; = \; 
L^{-d} \; \sum\limits_{\bf{k}} 
\; ( \xi^{-2} \; + \;{\bf{k}}^2  )^{-1} \; \quad - \quad
\int\limits_{\bf{k}} \; (\xi^{-2}\; + \; {\bf{k}}^2 )^{-1} \quad . 
\end{equation}
Here the range of ${\bf{k}}$ is limited by the cutoff $\Lambda$ 
according to $|k_j| \leq \Lambda$ for the bulk integral $\int\limits_{\bf{k}}$
and $- \Lambda \leq k_j < \Lambda$ for the sum $\sum_{\bf{k}}$. For $\Lambda
\to \infty$ the function $\tilde{D}_{field}$ becomes identical with
$\tilde{D}$ for $\tilde{a} \to 0$. Thus 
we decompose 
\begin{equation}
\tilde{D}_{field} \; = \; L^{2-d} \; 
\left[ I (L^2/\xi^2) \; + \; M_{field}  (\Lambda L, \; 
\Lambda \xi) \right] \; 
\end{equation}
with $M_{field} (\infty, \infty) = 0$ where the 
function $I(L^2/\xi^2)$ is the same as for the lattice model for 
$\tilde{a} \to 0$, (90) and (91),
but the cutoff dependent part $M_{field}$ of $\tilde{D}_{field}$ 
differs fundamentally
from the $\tilde{a}$ dependent part $M$ of $\tilde{D}$, as we shall see.

Application of the RG analysis of Section 3 to the field-theoretic model
leads to 
\begin{equation}
\frac{\chi_b - \chi}{\chi_b} \; = \; 4(n+2)u(\ell) (\xi/L)^{d-2}
\; I(L^2/\xi^2)
\Big[ 1 \; + \; R_{field}(L/\xi, \Lambda L) \Big] \; + \; O 
\left[ u(\ell)^2 \right]
\end{equation}
with
\begin{equation}
R_{field} (L/\xi, \Lambda L) \; = \; \frac{M_{field} \;  (\Lambda L, 
\Lambda \xi)}{I (L^2/\xi^2)} \quad .
\end{equation}
Equation (116) is valid for $d \leq$ 4. 
For $\Lambda L \gg 1$ and $\Lambda \xi \gg 1$ we have found 
\cite{15,ChDo3,21} 
\begin{equation}
M_{field} (\Lambda L, \Lambda \xi) \; = \; - d \; a_0 (d)(\Lambda L)^{d-4} \;
+ \; O \left[  (\Lambda L)^{d-6} , \exp (-\Lambda^{-2} \xi^{-2}) \right]
\end{equation}
where
\begin{equation}
a_0(d) \; = \; \frac{2\pi}{3} \; \int\limits_0^\infty
dx x \; e^{-x} \; 
\left[ \frac{1}{2\pi}\int\limits^1_{-1} dy \; \exp (- y^2 x) \right]^{d-1}
\quad .
\end{equation}
Together with $I(L^2/\xi^2)$, (101), this yields
the large $L \Lambda$ behavior of $R_{field}$ at fixed $\Lambda \xi \gg 1$
\begin{equation}
R_{field} (L/\xi, \Lambda L) \; = \; - (2 \pi)^{(d-1)/2}a_0 (d) 
(L/\xi)^{(3-d)/2} (\Lambda L)^{d-4} \; e^{L/\xi}.
\end{equation}
We see that, at fixed $\Lambda \xi \gg 1 \; , \;  R_{field}$ diverges 
exponentially towards $- \infty$ 
for $L/\xi \to \infty$  
and $\Delta \chi$ has the asymptotic size dependence in this limit (for cubic
geometry and $d < 4$)
\begin{equation}
\frac{\chi_b \; - \; \chi}{ \chi_b} \; \sim \; g(L/\xi) 
R_{field} (L/\xi, \Lambda L)\qquad \qquad \qquad
\end{equation}
\begin{equation}
 \; = \;  -  4(n+2) u^* d \; a_0 (d) (\Lambda \xi)^{d-2} (\Lambda L)^{-2} \; .
\end{equation} 
For the non-cubic geometries defined in Subsection 4.1 the corresponding 
results are obtained by substituting $I(L^2/\xi^2)$ in the form of (110)
instead of (101) and by the replacement $d \rightarrow \tilde{d}$ in the 
prefactor of (122). $R_{field}$ remains unchanged.

Like the result (106) for the lattice model, the behavior (122) is 
nonuniversal and
violates finite-size scaling, as pointed out recently \cite{15}.
The non-scaling effect becomes significant in the region below the
dashed line in Fig. 1 which, for the field-theoretic model, is determined by 
$|R_{field}| = 1$ \cite{15}, 
\begin{equation}
(2 \pi)^{(1-d)/2} (L/\xi)^{(d-3)/2} 
\exp (-L/\xi) \; = \; a_0(d) (\Lambda L)^{d-4} \;,
\end{equation} 
for both cubic and non-cubic geometries.

The structure of (122) differs fundamentally from that of (106) for the lattice
model in three respects: (i) the size-dependence of (122) is non-exponential, 
(ii) the dependence on $\Lambda$ is non-exponential, and (iii) the sign of 
(122) is negative. Since for $\xi \gg L$ we must have
$\chi_b - \chi > 0$ this sign implies the existence of a crossing point
of the bulk curve for $\chi_b$ and the curve for $\chi$ at some $T > T_c$, 
in contrast to (106) where $\chi$ does not cross the bulk
curve for $T \geq T_c$. We conclude that the $\varphi^4$ field theory does
not correctly describe the leading finite-size deviations from the bulk
critical behavior of lattice systems below four dimensions. This is true
also at $d \geq 4$; for $d = 4$
this follows from (97), (103) and (116), (120).
The case $d > 4$ will be discussed elsewhere \cite{CDnew}.

In Section 3 we have employed renormalization conditions in order to define the
renormalized theory at finite $\tilde{a}$ and finite $\Lambda$ for $d \leq 4$. 
For a calculation of the
universal part $g(L/\xi)$ of the finite-size effect, however, it is
possible to employ a more convenient RG approach using dimensional 
regularization and minimal subtraction at fixed $2 < d < 4$ \cite{SD,D85},
as has been done in the finite-size calculations of Refs. [39-41,46,47,50]. In
this case the fixed point value $u^*$ in (106) and (122) is replaced by 
$A^{-1}_d u^*_{min}$ where now $u^*_{min}$ can be taken from the accurate
Borel-resummed results of the minimally renormalized bulk theory 
\cite{SD,Russ,LMSD}.
For $d = 3$ the corresponding values are $A^{-1}_3 \; = \; 4 \pi$ and
$u^*_{min} \; = \; 0.0404,\;   0.0362, \;  0.0327$ for $n = 1,2,3,$ 
respectively \cite{LMSD}. In the large $n$-limit (at fixed $un$) the exact 
fixed point value is $u^*_{min} n = (4-d)/4$ for $ 2 < d < 4$.

\section{ Exact Results in the  Large-${\bf{n}}$ Limit}

In the following we perform the analysis of finite-size effects in the
$\varphi^4$ model above $T_c$ 
in the large-$n$ limit at finite lattice constant and finite cutoff for 
$d < 4$ without using the renormalization group.
The exact results in this limit will confirm the perturbative RG results
of the preceding Section, in particular  the
existence of a non-scaling region for both the field-theoretic and the
lattice $\varphi^4$ model in the range $L \gg  \xi$. The existence of this
region was overlooked in our recent work \cite{21}.

\subsection{Lattice Model at Finite Lattice Constant for 
$n \rightarrow \infty$}

We start from the exact result for the susceptibility $\chi/n = \hat{\chi}$
per component of the $\varphi^4$ lattice model  above $T_c$ in the limit $n \to \infty$ at fixed $u_0 n$ in a 
cubic geometry as determined 
by the implicit equation \cite{21} 
\begin{equation}
\hat{\chi}^{-1} \; = \; r_0 \; + \; 4 u_0 n L^{-d} \; \sum\limits_{{\bf{k}}}
(\hat{J}_{\bf{k}} \; + \; \hat{\chi})^{-1}\; .
\end{equation}
For $d > 2$ this can be rewritten as
\begin{equation}
\hat{\chi}^{-1} \; = \; r_0 - r_{0c} \; + \; 4 u_0 n D(\hat{\chi}^{-1}, L, 
\tilde{a}) \; 
- \; 4 u_0 n \hat{\chi}^{-1} \; \int\limits_{\bf{k}} \; 
\left[ \hat{J}_{\bf{k}} (\hat{J}_{\bf{k}} \; + \; \hat{\chi}^{-1})  
\right]^{-1}
\end{equation}
where $D (\hat{\chi}^{-1}, L, \tilde{a})$ is defined by (85) and
\begin{eqnarray}
r_{0c} &\; = \; & -4 u_0 n \; \int\limits_{\bf{k}} \; \hat{J}^{-2}_{\bf{k}} 
\; . 
\end{eqnarray}
The bulk susceptibility $\hat{\chi}_b$ and the bulk correlation length $\xi$ 
are determined by 
\begin{equation}
\hat{\chi}^{-1}_b \; = \; r_0 - r_{0c} \; - \; 4u_0 n \hat{\chi}_b^{-1} \;
\int\limits_{\bf{k}} \;  \left[ \hat{J}_{\bf{k}} \; (\hat{J}_{\bf{k}} \; + 
\; \hat{\chi}^{-1}_b) \right]^{-1} \quad , 
\end{equation}
\begin{equation}
\xi^2 \; = \; J_0 \hat{\chi}_b \quad .
\end{equation}

Although RG arguments will not be needed in the following we note that the
ultraviolet $(\tilde{a} \to 0)$ behavior of $\hat{\chi}$ , (125), is the 
same as that of $\hat{\chi}_b$ because the function $D$ has no ultraviolet 
($\tilde{a} \rightarrow 0$) divergence. This supports
the assumption made in Sect. 3. In the following we keep $\tilde{a}$ finite.

For $L \gg \tilde{a}$ and $\xi \gg \tilde{a}$,  i.e., for small 
$\hat{\chi}^{-1} \tilde{a}^2$ at finite
$\tilde{a}$, the bulk integral in (125) yields for $2 < d < 4$
\begin{equation}
\int\limits_{\bf{k}} \left[ \hat{J}_{\bf{k}} \; (\hat{J}_{\bf{k}} \; + 
\; \hat{\chi}^{-1}) \right]^{-1} \; = \; J^{-d/2}_0 \; A_d 
\; \hat{\chi}^{\epsilon/2}
\; \epsilon^{-1} 
\left\{ 1 \; + \; O \left[ ( \hat{\chi}^{-1} \tilde{a}^2)^{\epsilon/2} 
\right] \right\}
\end{equation}
with $\epsilon = 4 - d$ and 
\begin{equation}
A_d \; = \; 2^{2-d} \; \pi^{-d/2} (d-2)^{-1} \Gamma(3-d/2) \quad .
\end{equation}
For $u_0 n J_0^{-d/2}A_d \epsilon^{-1} \hat{\chi}^{\epsilon/2} \gg 1$ 
this leads to 
\begin{equation}
\hat{\chi} = \hat{\chi}_b
 \left[ 1 + \epsilon A^{-1}_d  \; \xi^{d-2} \tilde{D} (J^{1/2}_0 
\hat{\chi}^{1/2}, L, \tilde{a}) \right]^{2/(2-d)}
\end{equation}
where the function $\tilde{D}$ is defined by (88). For $L \gg \xi$ we
may replace $J^{1/2}_0 \hat{\chi}^{1/2}$ by  $J^{1/2}_0 \hat{\chi}^{1/2}_b \; 
= \;\xi$ in $\tilde{D}$. Using the decomposition (93) we arrive at
\begin{equation}
\Delta \hat{\chi} \; \equiv \; 
\frac{\hat{\chi}_b - \hat{\chi}}{\hat{\chi}_b}\; = \; 
\hat{g} (L/\xi) \left[ 1 + \hat{R} (L/\xi, \tilde{a}/\xi) \right] 
\end{equation}
where
\begin{equation}
\hat{g}(L/\xi) \; = \; 
2\epsilon A_d^{-1}(d-2)^{-1} (\xi/L)^{d-2} I (L^2/\xi^2) 
\end{equation}
with $I (L^2/\xi^2)$ given by (101), and
\begin{equation}
\hat{R}(L/\xi, \tilde{a}/\xi) \; = \; \exp 
\left[ \frac{1}{24} \Gamma (\tilde{a}/\xi) \frac{L}{\xi} \right] -1
\end{equation}
with $\Gamma (\tilde{a}/\xi)$ given by (104). The structure of $\hat{g}$
agrees with that of $g$, (100). (We note that the factor $\epsilon A_d^{-1}$
in (133) can be interpreted as $4u^* n$ where $u^*$ is the fixed point value
in the large-$n$ limit, see the last paragraph of Section 4.)
The function $\hat{R}$ turns out to be same as $R$, (103) and (104),
which was to be expected because $R$ is independent of $n$ (in one-loop 
order). 

For non-cubic geometries defined in Section 4.1 the result is only 
modified by the replacement $d \to \tilde{d}$ in the prefactor of the 
expression for $I(L^2/\xi^2)$ as given by (110), thus $\hat{g}$ reads 
explicitly
\begin{equation}
\hat{g} (L/\xi) = 2 \; \tilde{d} \; \pi^{1/2}\; [\Gamma(\epsilon/2)]^{-1}
(2\xi/L)^{(d-1)/2} e^{-L/\xi}.
\end{equation}

Equations (132)-(135) prove that finite-size scaling is violated for 
$2 < d < 4$ in the large-$n$ limit
for $L/\xi \gg$ at fixed $\tilde{a}/\xi > 0$ above $T_c$. This exact result
supports the correctness of our conclusions in the preceding Section based
on one-loop results. We note that the results (133)-(135) have finite limits
for $d \rightarrow 2$ at fixed $\xi$.

\subsection{ Field-Theoretic Model at Finite Cutoff for 
$n \rightarrow \infty$}

The analysis of the susceptibility $\hat{\chi} = \chi/n$ in the field-theoretic
$\varphi^4$ model at finite $\Lambda$ for $n \to \infty$ at fixed 
$u_0 n$ is parallel to that in
Section 5.1. Equations (124) - (130) remain valid after the replacements
$\hat{J}_{\bf{k}} \to {\bf{k}}^2$ and $J_0 \to 1$. 
Instead of (131) we now obtain for
$L \gg \Lambda^{-1}$ and $\xi \gg \Lambda^{-1}$ (for $2 < d < 4$) \cite{15}
\begin{equation}
\hat{\chi} \; = \; \hat{\chi}_b \; 
\left[ 1 \; + \; \epsilon A_d^{-1} \; \xi^{d-2} \; \tilde{D}_{field} 
(\xi, L, \Lambda) 
\right] ^{2/(2-d)}\; 
\end{equation}
where $\tilde{D}_{field}$ is defined by (114) and (115). For $L \gg \xi$ 
we arrive  at
\begin{equation}
\Delta \hat{\chi} \; \equiv \; \frac{\hat{\chi}_b - \hat{\chi}}{\hat{\chi}_b}
\; = \; \hat{g} (L/\xi)
\left[ 1 + \hat{R}_{field} (L/\xi, \Lambda L) \right]
\end{equation}
where $\hat{g}$ is identical with (133) or (135). Furthermore 
$\hat{R}_{field}$ is
identical with $R_{field}$, (120), as expected because $R_{field}$ is 
independent of $n$ (in one-loop order). These exact results confirm the
conclusions  drawn in Section 4 about the violation of finite-size scaling in
the $\varphi^4$ theory and about the failure of the continuum approximation
for finite lattice systems in the region $L \gg \xi$ above $T_c$.

\subsection{ Spherical Model}

Since  the $\varphi^4$ lattice model in the large-$n$
limit and  the spherical
model \cite{Joyce,Baxter} are expected to yield asymptotically identical 
results (for the case of periodic boundary conditions) we 
compare our results with those by Barber and Fisher 
\cite{17} and by Singh and Pathria
\cite{19}. Here we only study the approach
of the susceptibility towards the bulk critical behavior above $T_c$.

Barber and Fisher (BF) considered the spherical model with a film geometry
which should correspond to our result (132), (134) and (135) with 
$\tilde{d} = 1$. 
The result of BF for the dimensionless
susceptibility $\chi_{BF}$ reads for fixed $T > T_c$ and $L/\tilde{a} 
\to \infty$
\begin{equation}
\chi_{BF} \;  \sim  \;  \chi^b_{BF} \; + \;
J_0^{-1} B^0_d (T) (L/\tilde{a})^{(1-d)/2} \exp 
\left[ - \Gamma_d (T) L/\tilde{a} \right]  
\end{equation}
where $\chi^b_{BF}$ is the dimensionless bulk susceptibility and
$J_0 = 2 J$. (Here we have corrected a misprint in (8.11) of Ref.\cite{17} by 
replacing the exponent $(3-d)/4$ by $(1-d)/2$, compare (8.9) and (8.10) of 
Ref.\cite{17}.) Both functions $B^0_d(T)$ and $\Gamma_d(T)$ are expressed in
terms of the dimensionless function $\Phi_0(T)$ which is given by the
dimensionless inverse bulk susceptibility according to $\Phi_0(T) = J_0^{-1}
(\chi^b_{BF})^{-1}=(\tilde{a}/\xi)^2$. 
Calculating the small-$\Phi_0$ behavior of the derivative 
of the generalized Watson function $W_d (\Phi_0)$ as   
\begin{equation}
W^\prime_d (\Phi_0) \sim - 2^{-d} \pi^{-d/2} \; \Gamma (\epsilon/2) \; 
\Phi_0^{-\epsilon/2}
\end{equation} 
with $\epsilon = 4-d$ we have found
\begin{equation}
B^{(0)}_d (T) \; \sim \; - 2^{(d+1)/2} \pi^{1/2} 
\left[ \Gamma (\epsilon/2) \right]^{-1}\; \Phi_0^{-(d+3)/4}
\end{equation}
for $T \to T_c$. This leads to 
\begin{equation}
\frac{\chi_{BF}^b - \chi_{BF}}{\chi^b_{BF}} \; \sim 2 \;  
\pi^{1/2}\; [\Gamma(\epsilon/2)]^{-1}(2\xi/L)^{(d-1)/2} 
e^{-\Gamma_d (T)L/\tilde{a}}.
\end{equation}
The non-exponential part agrees with that of our $\hat{g}(L/\xi)$, (135), for
$\tilde{d}=1$.
If we expand the exponent 
$\Gamma_d(T) = 2 \;  {\mbox{arcsinh}} (\Phi_0^{1/2} /2) \;$
to {\it{third}} order
in $\Phi^{1/2}_0$ and express it in terms of $\tilde{a}/\xi$ we obtain
\begin{eqnarray}
\exp \bigg[ - \Gamma_d (T) L/\tilde{a} \bigg] \; = \; 
\exp \bigg[ - (2L/\tilde{a})\; {\mbox{arcsinh}}\; \bigg( \frac{1}{2} 
\tilde{a}/\xi 
\bigg) \bigg]\\[1cm]
\; = \; 
\; \exp 
\bigg\{ -L/\xi  \; +  \; 
\frac{1}{24} (\tilde{a}/\xi)^2 \; L/\xi 
\; + \; O  \left[    (\tilde{a}/\xi)^4 L/\xi \right]
\bigg\}
\end{eqnarray}
which also agrees with the exponential part of our solution (132), (134), 
(135) of 
the $\varphi^4$ lattice model. The importance of the {\it positive} second
term $\propto (\tilde{a}/\xi)^2$
in (143) for the leading finite-size deviations from bulk critical behavior
(for $L/\xi \to \infty$ at fixed $\tilde{a}/\xi > 0$) was overlooked by BF 
\cite{17} (who considered the exponent in (138) only in the limit 
$T \to T_c$ at fixed $L/\tilde{a} < \infty$ rather than 
$L/\tilde{a} \to \infty $ at fixed $T > T_c$, see also equations (5.75)-(5.77)
of Ref.\cite{1}).

The solution of Singh and Pathria (SP) \cite{19} for
$\Delta \hat{\chi}$ (see equation (19) of Ref. \cite{19}) in the large-L
limit at fixed $T > T_c$ agrees with the universal part $\hat{g}$ of 
our result, (135). 
Our non-universal contribution $\hat{R}$, (134), is not contained in the
solution of SP.

\section{Summary and Conclusions}

In the following we summarize and further comment on the results
of this paper as follows.

We have studied the consequences of renormalizability (in terms of bulk 
renormalizations) of the
$\varphi^4$ theory in a confined geometry with periodic boundary
conditions below four dimensions. We have found that the consequences
for the confined system are in contrast to those for the bulk system.
While for the bulk system renormalizability implies that the leading
critical {\it{temperature}} dependence is not significantly affected
by a finite lattice constant $\tilde{a}$ or by a finite cutoff $\Lambda$
this is not generally the case for the leading {\it{size}}
dependence of the confined system. Lattice and cutoff effects are
asymptotically negligible as criticality is approached at fixed finite
ratio $L/\xi$ but not in the limit $L/\xi \gg 1$ above $T_c$.
In the latter case the {\it leading} finite-size effect on the susceptibility 
$\chi$ turns out to be 
nonuniversal, i.e., to depend explicitly on $\tilde{a}$ and $\Lambda$ 
[see equations (106) and (122)],
and to violate finite-size scaling in the region $L \gg \xi$ of the 
$L^{-1} - \xi^{-1}$ plane (Fig. 1).

Although lattice and continuum models yield the same asymptotic critical 
behavior of bulk systems we have shown that this is not the case for
confined systems. While lattice systems (with periodic boundary conditions) 
have an exponential size
dependence of $\Delta \chi$ above $T_c$ for large $L/\xi$, a power-law 
behavior $\Delta \chi \propto (\Lambda L)^{-2}$ is obtained from the 
$\varphi^4$ field theory \cite{15}.

It is expected that part of  our conclusions apply also to 
$T < T_c$ for the case $n = 1$.
For $n \geq 2 $ (and for periodic boundary conditions) the power-law 
behavior of the Goldstone modes is expected to govern
the finite-size deviations from bulk critical behavior such that nonuniversal
exponential terms become subleading .

Our results for the leading finite-size deviations  from bulk critical
behavior have been derived for $2 < d \leq 4$ dimensions but our conclusions 
may be applicable to $d = 2$ dimensions. A detailed reexamination of the 
exponential terms in the existing exact results for the two-dimensional Ising 
model \cite{16,AY} would be interesting.

Part of our results remain valid also at and above four dimensions. One 
of the consequences is that the universal two-variable finite-size 
scaling form
for the $\varphi^4$ lattice model for $d > 4$ [59,60,63,66-68] is violated
for $L/\xi \gg 1$ above $T_c$ for general $n$ and below $T_c$ for $n = 1$.
A further consequence is that the predictions of the lowest-mode approach
\cite{7} and of the phenomenological scaling theory implying a single-variable
scaling form \cite{BNPY,KB85} fail qualitatively in the region $\xi/L \gg 1$
where these theories predict a universal power-law behavior 
$\Delta \chi \propto L^{-d}$
instead of a non-universal exponential size dependence $\Delta \chi \propto 
e^{-c L}$. The latter can easily be incorporated
in our theory above four dimensions \cite{ChDo3} by extending our present
perturbation approach of Section 4 to $d > 4$ \cite{CDnew}.

From a quantitative point of view, our prediction of a violation of finite-size
scaling is difficult to be tested by means of Monte Carlo simulations 
(e.g., for Ising models) because the non-scaling effect on $\chi$ occurs 
predominantly in the region $L \gg \xi$ where the total finite-size effects 
on $\chi$ are exponentially small (for periodic boundary conditions).

Our general arguments regarding the consequences of renormalizability
(as presented in Section 3) are presumably not restricted to periodic 
boundary conditions
but may be  generalized to {\it non-periodic} boundary conditions. 
We consider this to be potentially
important for applications to real systems where finite-size deviations from
bulk critical behavior are not exponentially small. We do no longer see 
a stringent
reason to believe that renormalizability implies the validity of finite-size
scaling in the more complicated cases of non-periodic boundary conditions 
where additional renormalizations and nonuniversal length scales come into
play.
In particular for the important case of confined $^4$He
near the superfluid transition where the entire region
$L > \xi$ and $L \leq \xi$ is perfectly well accessible to
high-resolution experiments [69,70,72-77,79,80,90] 
we cannot exclude the existence of
nonuniversal non-scaling effects in the $\varphi^4$ (lattice and field)
theory with Dirichlet boundary conditions. This could eventually lead
to a natural explanation of longstanding and recent discrepancies
between experimental data [69-80] and theoretical  
predictions [5,11,24,29,31,32,36,37,42,49]
that were based on (seemingly plausible) assumptions which imply 
the validity of finite-size scaling.
Also for the exploration of finite-size effects on transport properties
in $^4$He on earth [72,74,79,90] and under microgravity conditions 
\cite{Ahlers} as well as on thermodynamic properties near  
ordinary critical points under microgravity conditions \cite{Bar},
detailed knowledge on the effect of a finite atomic distance may turn out to
be important.

{\bf{Acknowledgments}}

Support by Sonderforschungsbereich 341 der Deutschen 
Forschungsgemeinschaft and by NASA under contract numbers 960838 and
100G7E094 is acknowledged. One of us (X.S.C.) thanks the National Science 
Foundation of China for support under Grant No. 19704005.

\newpage

\section*{Appendix A: Bulk Correlation Length at Finite Lattice Constant}

\renewcommand{\theequation}{A.\arabic{equation}}
\setcounter{equation}{0}

In this Appendix we determine the bulk correlation length $\xi$ 
above $T_c$ as a
function of $t$ at finite lattice spacing for $d \leq 4$.  
The derivation is parallel to
that at infinite cutoff in Ref. \cite{SD}. 
We introduce the renormalized
temperature variable
\begin{equation}
r \; = \; Z^{-1}_r \; (r_0 - r_{0c})= Z_r^{-1} a_0 t
\end{equation}
where $Z_r$ is identical with the $Z$ factor $Z_{\varphi^2} = Z_r$
that is needed to renormalize the bare vertex function $\tilde{\Gamma}^{(1,2)}
 \; (\xi, u_0, \tilde{a}, d)$ \cite{14,Amit,13} for $d \leq 4$
\begin{equation}
\tilde{\Gamma}_R^{(1,2)}(\xi, u, \mu, \tilde{a}, d) \; = \; Z_r \;
Z_\varphi \; \tilde{\Gamma}^{(1,2)} \; 
(\xi, \mu^\epsilon \; J^2_0 \; Z_u \; u, \tilde{a}, d)
\quad .
\end{equation}
The one-loop expression for 
$\tilde{\Gamma}^{(1,2)}$ is
\begin{equation}
\tilde{\Gamma}^{(1,2)} \; (\xi, u_0, \tilde{a}, d)
\; = \; 1 \; - \; 4(n+2) \; u_0 \int\limits_{\bf{k}} 
\left[ \hat{J}_{\bf{k}} + J_0 \; \xi^{-2}  \right]^{-2} \; + \; O(u^2_0).
\end{equation}
The $Z$ factor $Z_r (u, \mu \tilde{a}, d)$ for $d \leq 4$ is determined
by the renormalization condition at $\xi \;  = \;  \mu^{-1}$
\begin{equation}
\tilde{\Gamma}^{(1,2)}_R \;(\mu^{-1}, u, \mu, \tilde{a}, d) \; = \; 1 
\end{equation}
which yields in one-loop order
\begin{equation}
Z_r (u, \mu \tilde{a}, d) \; = \; 1 + 4(n+2)  \; u I(\mu \tilde{a}, d)
\; + \; O(u^2)
\end{equation}
where $I (\mu \tilde{a}, d)$ is given in (35).
Using (A.1) we rewrite the right-hand side of (17) in terms of renormalized
quantities as
\begin{eqnarray}
r_0 - r_{0c} &\; = \;& h(\xi, \mu^{\epsilon} J^2_0 \; Z_u 
u, \tilde{a}, d)\\[1cm]
&\; = \;& Z_r (u, \mu \tilde{a}, d) \; \mu^2 Q (\mu \xi, u, \mu \tilde{a}, d)
\end{eqnarray}
with the dimensionless amplitude function $Q$. Taking the derivative at fixed
$u_0, \tilde{a}$ and $r_0 - r_{0c}$ (i.e. fixed $\xi$) yields the RGE
\begin{eqnarray}
\left[ \mu \partial_\mu \: + \; \beta_u \partial_u \right.  
&\; + \;& \left. (2-\zeta_r) \right]\;  Q \;  (\mu \xi , u, \mu
\tilde{a}, d) \; = \; 0  \quad ,\\[1cm]
\zeta_r (u, \mu \tilde{a}, d) &\; = \;& (\mu\;  \partial_\mu \; \ln\;  
Z_r^{-1})_0 \quad .
\end{eqnarray}
The formal solution is 
\begin{equation}
Q (\mu \xi, u, \mu \tilde{a}, d) \; = \; Q(1, u(\ell), \tilde{a}/\xi, d)
\exp \; \int\limits_1^\ell \; \left(\zeta^*_r - \zeta_r (\ell^{\prime}) \right]
\frac{d \ell^{\prime}}{\ell^{\prime}} \quad 
\end{equation}
with $\ell \; = \; (\mu \xi)^{-1}$ where $\zeta^*_r \; = \; 
\zeta_r (u^*, 0, d)$
and $\zeta_r (\ell) \; \equiv \; \zeta_r
(u(\ell), \ell \mu \tilde{a}, d)$. Equations (A.1), (A.7) and (A.10) can be
summarized as
\begin{equation}
r \; = \; at \; = \; \xi^{-2} Q (1, u(\ell), \tilde{a}/\xi, d) \; \exp \;
\int\limits^{1/\mu \xi}_1 \zeta_r (\ell^{\prime}) \; 
\frac{d \ell^{\prime}}{\ell^{\prime}}
\end{equation}
with
\begin{equation}
a \; = \; Z_r (u, \mu \tilde{a}, d)^{-1} \; a_0 \; > \; 0
\end{equation}
where the bare parameter $a_0$ is defined in (11). 
Equation (A.11) determines $t > 0$ as a function of $\xi$ for $d \leq 4$ at
finite $\tilde{a}$. 
Inverting (A.11)
yields $\xi(t)$ including non-asymptotic (Wegner \cite{Wegner}) corrections. 
The asymptotic $(\xi \to \infty)$ form of the correlation
length follows from (A.11) as
\begin{equation}
\xi \; = \; \xi_0 \; t^{-\nu}
\end{equation}
with the critical exponent
\begin{equation}
\nu \; = \; (2 - \zeta^*_r)^{-1}
\end{equation}
and the amplitude
\begin{equation}
\xi_0 \; = \; \mu^{2\nu-1}\; a^{-\nu} \;
\left\{ Q^* \exp \; \int\limits^0_1 \; \left[ \zeta^*_r \; - \zeta_r \; 
(\ell^{\prime}) \right] \; \frac{d \ell^{\prime}}{\ell^{\prime}} \right\}^{\nu}
\end{equation}
where
\begin{eqnarray}
Q^*   \; = \;  Q (1, u^* , 0, d) \quad .
\end{eqnarray}
After the choice $\mu = \xi^{-1}_0$, the correlation-length amplitude
$\xi_0$ is determined implicitly in terms of the bare parameter $a_0$, 
the lattice spacing $\tilde{a}$ and the renormalized 
coupling $u$ by
\begin{equation}
\xi^2_0 \; = \; Z_r (u, \tilde{a}/\xi_0, d) \; a^{-1}_0 \; Q^* \exp \; 
\int\limits^0_1 \;
\left[ \zeta^*_r  \; - \; \zeta_r (\ell^{\prime}) \; \right] \;     
\frac{d \ell^{\prime}}{\ell^{\prime}}  \quad . 
\end{equation}
We note that the functions $h (\xi, u_0, \tilde{a}, d)$ and $Q(\mu \xi, u, \mu
\tilde{a}, d)$ do not have an expansion in integer powers of $u_0$ and $u$,
respectively, beyond one-loop order \cite{SD}. The one-loop expression of 
$Q (\mu \xi, u, \mu \tilde{a}, d)$ and of $Q^*$ can
be derived from (19), (34), (A.5) and (A.7). An integral representation of $Q$
in terms of expandable functions can be derived at finite $\tilde{a}$ along
the lines of Section 4.1 of Ref. \cite{SD}.

\newpage

{\bf{Figure Caption}}

{\bf{Fig. 1.}} 
Asymptotic $L^{-1} - \xi^{-1}$ plane in units of the lattice constant
$\tilde{a}$ above $T_c$ (schematic plot) for the $\varphi^4$ lattice model
below four dimensions where $L$ is the system size and $\xi$ is the bulk
correlation length. Nonuniversal lattice effects become nonnegligible in
the non-scaling region below the dashed line where
finite-size scaling is violated. This crossover line is determined by
equation (107). Well above this line the dependence on the 
lattice spacing $\tilde{a}$ is negligible in equation (3). The arrow
indicates an approach towards bulk critical behavior at constant $0 < t \ll 1$
through the non-scaling region where equation (106) is  valid.
A corresponding non-scaling region exists also for the field-theoretic 
$\varphi^4$  model at finite cutoff where the crossover line is determined by 
equation (123) and the non-scaling effect is described by equation (122),
compare Fig.1 of Ref.[52]. A corresponding crossover line should be added to 
Fig.1 of Ref.[63].

\end{document}